# Radiation-driven winds of hot luminous stars

## XVIII. The reliability of stellar and wind parameter determinations from spectral analysis of selected central stars of planetary nebulae and the possibility of single-star supernova Ia progenitors

C. B. Kaschinski, A. W. A. Pauldrach, and T. L. Hoffmann

Universitäts-Sternwarte München, Scheinerstraße 1, 81679 München, Germany
e-mail: corni@usm.lmu.de, uh10107@usm.lmu.de, hoffmann@usm.lmu.de

**ABSTRACT**

*Context.* The uncertainty in the degree to which radiation-driven winds of hot stars might be affected by small inhomogeneities in the density leads to a corresponding uncertainty in the determination of the atmospheric mass loss rates from the strength of optical recombination lines and – since the mass loss rate is not a free parameter but a function of the stellar parameters mass, radius, luminosity, and abundances – in principle also in the determination of these stellar parameters. Furthermore, the optical recombination lines also react sensitively to even small changes in the density structure resulting from the (often assumed instead of computed) velocity law of the outflow. This raises the question of how reliable the parameter determinations from such lines are.
*Aims.* The currently existing severe discrepancy between CSPN (central stars of planetary nebulae) stellar and wind parameters derived from model fits to the optical spectra and those derived using hydrodynamically consistent model fits to the UV spectra is to be reassessed via a simultaneous optical/UV analysis using a state-of-the-art model atmosphere code.
*Methods.* We have modified our hydrodynamically consistent model atmosphere code with an implementation of the usual ad-hoc treatment of clumping (small inhomogeneities in the density) in the wind. This allows us to re-evaluate, with respect to their influence on the appearance of the UV spectra and their compatibility with the observations, the parameters determined in an earlier study that had employed clumping in its models to achieve a fit to the observed optical spectra.
*Results.* The discrepancy between the optical and the UV analyses is confirmed to be the result of a missing consistency between stellar and wind parameters in the optical analysis. While clumping in the wind does significantly increase the emission in the optical hydrogen and helium recombination lines, the influence of the density (velocity field) is of the same order as that of moderate clumping factors. Moderate clumping factors leave the UV spectra mostly unaffected, indicating that the influence on the ionization balance, and thus on the radiative acceleration, is small. Instead of the erratic behavior of the clumping factors claimed from the optical analyses, our analysis based on the velocity field computed from radiative driving yields similar clumping factors for all CSPNs, with a typical value of $f_{\rm cl} = 4$. With and without clumping, wind strengths and terminal velocities consistent with the stellar parameters from the optical analysis give spectra incompatible with both optical and UV observations, whereas a model that consistently implements the physics of radiation driven winds achieves a good fit to both the optical and UV observations with a proper choice of stellar parameters. The shock temperatures and the ratios of X-ray to bolometric luminosity required to reproduce the highly ionized O VI line in the FUSE spectral range agree with those known from massive O stars ($L_{\rm X}/L_{\rm bol} \sim 10^{-7} \ldots 10^{-6}$), again confirming the similarity of O-type CSPN and massive O star atmospheres and further strengthening the claim that both have identical wind driving mechanisms.
*Conclusions.* The similarity of the winds of O-type CSPNs and those of massive O stars justifies using the same methods based on the dynamics of radiation-driven winds in their analysis, thus supporting the earlier result that several of the CSPNs in the sample have near-Chandrasekhar-limit masses and may thus be possible single-star progenitors of type Ia supernovae.

## 1. Introduction

The computation of a predicted (i.e., synthetic) spectrum that can be compared to the observed spectrum is the primary and most important diagnostic tool for determining the fundamental parameters of a star. But while a photospheric model[1] will yield information about the effective temperature $T_{\rm eff}$ and the surface gravity $\log g$, the spectroscopic determination of the radius $R$ requires an observable quantity that can be spectroscopically traced over a depth range covering an appreciable fraction of the physical size of the star. Furthermore, a physical description is required that predicts the run of this quantity as a function of the radius.

Many stars have thin, effectively plane-parallel atmospheres, for which such a procedure is not applicable. For these stars, additional information beside the stellar spectrum, for example the measured flux and the distance to the star, is required to determine the radius. In the atmospheres of O-type stars, however, the radiative intensities are large enough for the radiation pressure to drive a stellar "wind", propelled by the transfer of photon momentum to ions in the wind via absorption in spectral lines. The winds leave conspicuous, easily observable signatures in the stellar spectra, allowing individual spectral lines to be traced out to tens of stellar radii, thereby yielding information on the run of the velocity and density field in the wind. Because the density and velocity in the wind can also be predicted by selfconsistent hydrodynamical modelling of the outflow on the basis of the gravitational and the (computed) radiative acceleration at each point in the wind, the comparison of observed and predicted spectra provides a link to the stellar radius and gravity, thus allowing for a purely spectroscopic determination of the stellar mass.

[1] In classical photospheric models, the effective temperature $T_{\rm eff}$ is usually determined via the ionization balance, derived from the strengths of lines from successive ionization stages, and the surface gravity $\log g$ is determined via the density-dependent (and thus, pressure-dependent) Stark-broadening in the hydrogen Balmer line wings.

Central stars of planetary nebulae (CSPNs) are valuable research objects because they represent a unique phase in the evolution of intermediate-mass stars, and accurately known stellar parameters would constitute a stringent test of evolutionary models. Unfortunately, accurate *distances* to most CSPNs are not available, and information about their sizes (and thus, via $\log g$, their masses) must therefore be derived either spectroscopically through hydrodynamic modelling of the expanding atmosphere or, supplemental to spectroscopy, with the aid of additional physical considerations and models.

The current dispute regarding the true stellar parameters of CSPNs has its origins in two series of analyses of (largely congruent) samples of O-type CSPNs, using different methods. One of these (Méndez et al. 1988b,a; Kudritzki et al. 1997; Kudritzki et al. 2006) used model fits to the optical hydrogen and helium lines to determine effective temperatures and surface gravities. Lacking hydrodynamical modelling of the wind, they had to use stellar evolutionary models to determine masses and radii, comparing effective temperatures and surface gravities to the theoretical evolutionary tracks in the $T_{\rm eff}$–$\log g$ plane. These analyses were therefore not an independent test of the evolutionary models, since their validity had been assumed all along. Nevertheless, a surprising result was that Kudritzki et al. (1997) found many high-mass CSPNs (with masses around $0.9\,M_{\odot}$), which was unexpected, since theory predicted them to evolve too quickly to be observed in larger numbers. Still, the mass loss rates derived from their stellar parameters and from modelling of the H$\alpha$ emission were well in agreement with what was expected from radiation-driven wind theory, providing strong evidence that the winds of these objects are indeed driven by radiation pressure.

The other series of analyses (Pauldrach et al. 1988; Pauldrach et al. 2003, 2004; Kaschinski et al. 2012) is based on hydrodynamic modelling of the wind, using fits to the observable UV spectra. This type of analysis had been successfully applied to massive O stars (Pauldrach 1987; Pauldrach et al. 1990a; Pauldrach et al. 2001; Pauldrach et al. 2012), but its application to O-type CSPNs gave even more unexpected results: Pauldrach et al. (2004) found a number of CSPNs with masses around 1.3 to $1.4\,M_{\odot}$, near the Chandrasekhar limit for white dwarfs (cf. Chandrasekhar 1931). These high masses, however, were not simply due to a systematic overestimate of the masses by the method. One of the stars, NGC 2392, was even found to have a surprisingly small mass of only $0.4\,M_{\odot}$, compared to the $0.9\,M_{\odot}$ that had been derived by Kudritzki et al. (1997).

Since both sets of analyses were based on different spectral regions it was not clear whether this discrepancy was due to some assumption in the modelling that was perhaps unjustified (e.g., that the winds were radiatively driven, or that the mass–luminosity relation assumed by Kudritzki et al. (1997) correctly described these objects), or whether there was something peculiar about either the optical or the UV spectra, despite the good fits both groups achieved for their spectral ranges. A first attempt to resolve the issue was made by Kaschinski et al. (2012) with combined optical and UV modelling from the same atmospheric models. For their analysis they chose NGC 2392, the least massive star, and NGC 6826, the most massive star in the sample of Pauldrach et al. (2004). These extreme cases were chosen because covering a larger spread in stellar parameters makes it more likely for the results to show real physical behavior, and not just a selection effect, in a group of stellar objects.

The results of this investigation were astounding. The synthetic spectra modeled with the stellar parameters from the optical analysis could only achieve fits to the observations because the wind parameters that had been assumed in that analysis were inconsistent with the stellar parameters. Enforcing consistency between stellar and wind parameters could yield simultaneous fits to both the UV and the optical spectra, but only with stellar parameters that were in disagreement with the theoretical CSPN mass–luminosity relation. Merely the optical emission lines H$\alpha$ and He II $\lambda$4686 could not be reproduced fully by Kaschinski et al., the emission predicted by their models being too weak.

In the present paper we will address this significant point. For the optical analyses, H$\alpha$ is the primary diagnostic line for deriving mass loss rates, and reliable modelling of the H$\alpha$ emission is therefore crucial to obtain accurate mass loss rates in these analyses.[2] In recent years[3] evidence has accumulated that the winds might not be entirely smooth but clumpy (e.g., Repolust et al. 2004), and H$\alpha$, being a recombination line and thus proportional to the square of the density, would therefore show increased emission compared to that from a smooth wind. In other words, model fits assuming a smooth flow would overestimate the mass loss rates. In view of this fact, Kudritzki et al. (2006) reanalyzed the data of Kudritzki et al. (1997), adjusting clumping factors in the models to provide new fits to the optical emission lines and resulting in revised stellar parameters and significantly reduced mass loss rates. We will compare the results from our combined optical and UV analysis, now also taking clumping into account, to these revised parameters.

Further information about the nature of O-type CSPN winds is provided by X-ray observations (Guerrero et al. 2000; Kastner et al. 2012) that show central stars of planetary nebulae as X-ray emitters. This radiation is thought to originate, as in massive O star winds, from small packets of the gas that are accelerated[4] relative to the mean flow and collide with other gas fragments further out, resulting in small shocked regions (with a volume filling factor of around $f \simeq 10^{-2}$) of hot gas ($T \simeq 10^{6}$ K) that cools radiatively, emitting radiation in the X-ray regime (Lucy & White 1980; Lucy 1982). Apart from direct observation these X-rays can also be inferred from lines of highly ionized species (N V, O VI, and S VI) in the observable UV region. The presence of such highly ionized species (so-called "superionization", Snow & Morton 1976) could be readily explained by shocks being the source of the high-energy radiation required to generate these ions. Some of their lines lie in the FUSE spectral range, and corresponding observations exist for several of the CSPNs in our sample. We use these observations to ascertain whether

[2] The UV analyses are not so dependent on a single line, as an entire forest exists of lines (mainly Fe and Ni in ionization stages IV and V) that are formed at the base of the wind and react sensitively to the mass loss rate.

[3] It has long been known that the line-driving mechanism of the winds can be intrinsically unstable (Milne 1926; Lucy & Solomon 1970), but it was not suspected that the fragmentation of the wind would already be significant in the lower regions where most of H$\alpha$ is formed.

[4] Since the line force is proportional to the velocity *gradient*, a small localized disturbance in the velocity tends to be magnified in a runaway fashion.

CSPN winds differ from massive O star winds with regard to shock radiation.

In the following we will briefly review the pros and cons of the two spectral analysis techniques (UV and optical) (Sect. 2) and discuss the FUSE observations and our modelling of the shocks to provide constraints on the radiation leading to the production of O VI (Sects. 3 and 4). Then we will describe our implementation of clumping, providing reference comparisons to FASTWIND, the modelling code that had been used by Kudritzki et al. (2006), and we show comparisons to the observations of the synthetic spectra from our model runs, using the parameters of both Kudritzki et al. (2006) and Pauldrach et al. (2004), including and excluding clumping (Sect. 5). Finally, we discuss the implications of our findings in Sect. 6.

## 2. Methods for the investigation of CSPN winds and the determination of stellar parameters

A review of the current literature shows that at present there are two major approaches to the spectral analysis of hot stars with expanding atmospheres. The first treats the determination of the structure of the atmosphere as a problem in hydrodynamics, which can be solved using basic physics once the state of the gas and the radiation field is modeled in sufficient detail to enable the computation of the forces acting on each volume element of the atmosphere. The second treats the determination of the atmospheric structure as an optimization problem by parametrizing the velocity law and attempting to find the best solution by varying the parameters and choosing the set that results in the best fit to the observed spectral line profiles.

### 2.1. Details of the methods

**Parametrized method to fit the spectra of CSPNs.** This method has traditionally been the domain of optical spectral analysis, but it can be extended to include the UV as well (e.g., Herald & Bianchi 2011). In the analysis of the optical spectrum, the effective temperature $T_{\text{eff}}$ is obtained from NLTE model fits to the optical He I and He II stellar absorption lines and the surface gravity $\log g$ is derived from fits to the Balmer lines (mainly H$\gamma$, but also H$\beta$, cf. Méndez et al. 1988b, Kudritzki et al. 1997). The velocity of the outflow in the wind region is parametrized with a so-called "beta velocity law," $v(r) = v_\infty (1 - bR/r)^\beta$, where the best-fit value of $\beta$ is to be found as part of the analysis, and the parameter $b$ fixes the velocity at the inner boundary of the wind which is usually taken to be of the order of the sound speed. The terminal wind velocity $v_\infty$ can usually be measured directly from the observed blue edge of the saturated P-Cygni profiles in the UV spectrum. (If a line does not remain saturated until the terminal velocity is reached, its blue edge will of course indicate a smaller velocity.) This velocity law is then connected in a smooth transition to the hydrostatic density structure in the deeper atmospheric layers.

The mass loss rate $\dot{M}$ that determines the wind density via the equation of continuity, $\rho(r) = \dot{M}/(4\pi r^2 v(r))$, is determined via a model fit to the H$\alpha$ emission profile (cf. Kudritzki et al. 2006). Note, however, that this fitting procedure is actually only sensitive to the combined quantity $Q \sim \dot{M}(Rv_\infty)^{-3/2}$, and the determination of absolute mass loss rates depends on knowledge of the stellar radii $R$ which can not be derived with this analysis and must be provided by other means.

The realization that the winds may not be entirely smooth has given a provisional explanation to the fact that with a model assuming a unclumped flow it was not always possible to simultaneously fit all emission lines in the optical, and thus some form of ad-hoc treatment of clumping (such as the one described in Sect. 5) is nowadays often used with this method to improve the fit to the spectrum.

**Consistent hydrodynamic method for determining complete sets of stellar parameters for CSPNs.** Taking into account the relevant forces and actually solving the equation of motion eliminates the need to assume a velocity law. The solution not only provides the structure of the velocity field and the terminal velocity, but also the density (and thus, the mass loss rate) in a unified way throughout the entire atmosphere, from the hydrostatic layers deep in the photosphere to the outer regions of the wind where the forces become negligible and the terminal velocity is reached.

A complete hydrodynamically self-consistent model usually requires some iteration scheme, beginning with a plausible guess of the line forces for a given set of stellar parameters. Using this initial guess, the hydrodynamical structure is computed and the NLTE problem (i.e., the solution of radiative transfer, rate equations, and corresponding occupation numbers) is solved. From this solution, new values for the radiative forces are calculated, and the procedure is iterated until convergence is achieved. The radiative force is computed analogously to the synthetic spectrum using the same occupation numbers and radiation field, and the quality of the fit of the spectrum is thus also a measure of the accuracy of the computed line force. If the spectral fit of the converged model is not satisfactory, new stellar parameters are chosen and the process is repeated.

A distinctive characteristic of the method is that the mass loss rate and the terminal velocity that satisfy the constraint of self-consistency are determined in dependence of the stellar parameters. Since the UV spectrum varies characteristically as a function of these wind parameters (see, for example, Fig. 5 of Pauldrach et al. 2012), this dependence can be used diagnostically to determine the stellar parameters from the appearance of the UV spectrum. (The UV is much better suited for this analysis, since it contains numerous lines of varying strength formed at different depths throughout the atmosphere, whereas the lines in the optical are rather few and rather weak – cf. Fig. 1.) For a more detailed discussion of the numerical procedures used to implement this method we refer the reader to Pauldrach et al. 2012 and Pauldrach et al. 2001.

### 2.2. Discussion

The general impression in the community appears to be that both methods are of similar quality – one may be slightly better than the other, but both can be seen to be on equal footing. Although such a view may be "politically correct", objectively it is a wrong assessment of the situation. With regard to a purely spectroscopic determination of a complete set of stellar parameters *the self-consistent hydrodynamical method has no alternative*. It is not only superior, but is in fact essential for this task.

To understand why this is so, consider that the hydrodynamic method has as its basis the *physics* of the problem, and while perhaps not perfect, at least attempts to describe the outflow on the basis of well-known principles – and does this successfully. Most importantly, it predicts the relation between the properties of the wind (and thus, the appearance of the observable spectrum) and the stellar parameters and thereby allows the determination of a complete set of stellar parameters from the (UV) spectrum alone.

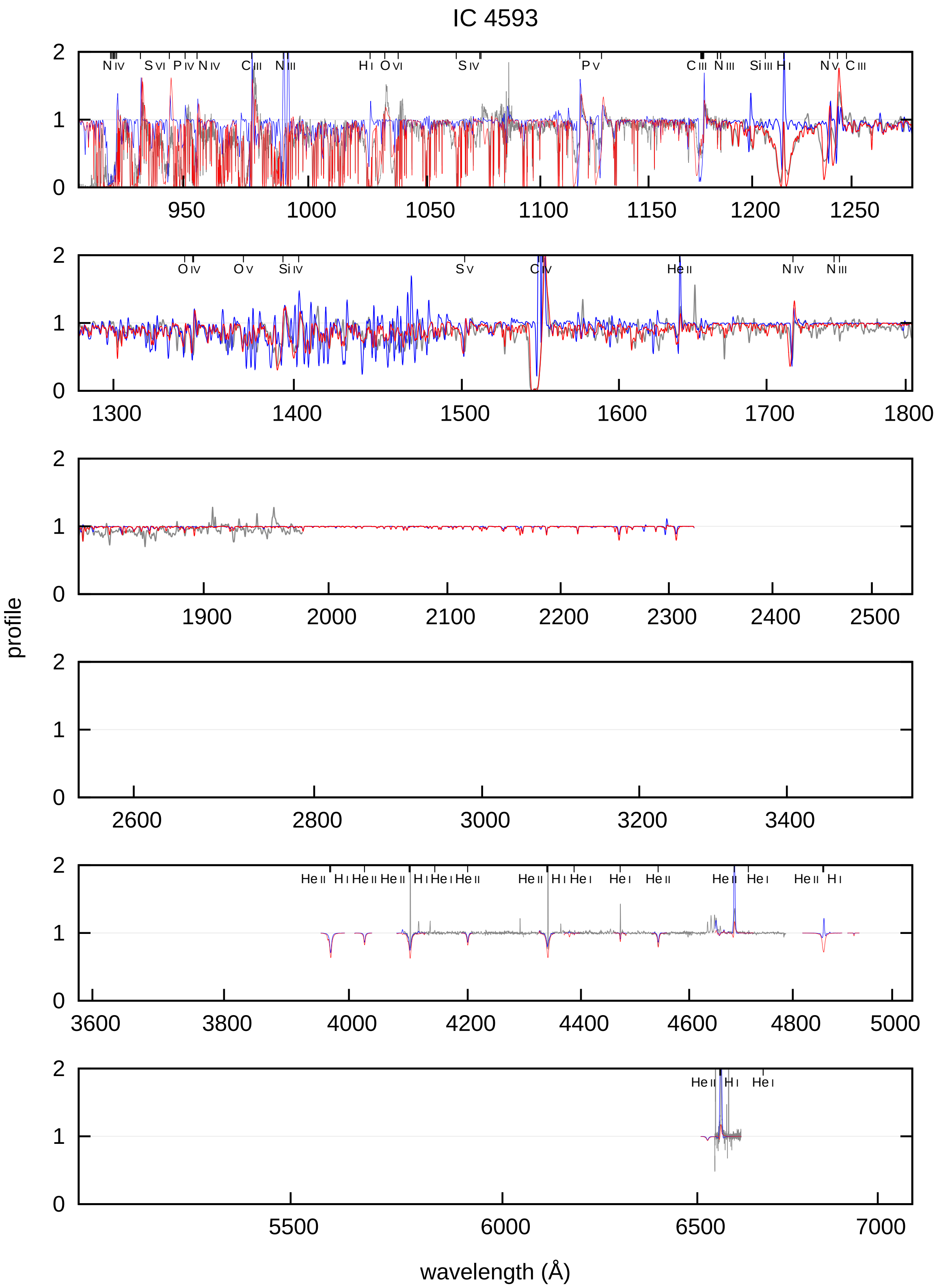

**Fig. 1.** Observed (gray) UV and optical spectra for IC 4593 compared to consistently calculated synthetic spectra based on the stellar parameter sets of Kudritzki et al. (2006) (K06c; blue) and Pauldrach et al. (2004) (P04; red). As the K06c spectrum is clearly incompatible with the observations, this parameter set has to be ruled out, whereas the P04 spectrum is generally in good agreement with the observations (here the most important interstellar lines have been included in the FUSE spectral part), offering a set of stellar parameters which is also consistent with the picture of the X-ray, the wind-dynamical properties (cf. Kaschinski et al. 2012), and the wind-momentum-luminosity relation. Although the *spectrum in the optical wavelength range* is also computed by our code, it *is not our means of spectral diagnostics*, because compared to the UV it contains only few and only rather weak lines which (in particular H$\alpha$ and He II $\lambda$4686) moreover may be strongly influenced by second-order effects possibly connected to small inhomogeneities in the density. With regard to possible discrepancies between observed and predicted spectra there is no reason to assume that for modeling O-type atmospheres the optical range carries the same diagnostic weight as the UV: when compared on the same vertical scale, as shown here, it becomes immediately obvious that the UV spectral range contains much more information, and due to the much stronger lines in the UV the totality of features is far less sensitive to minor uncertainties in the modelling than in the optical range. This makes the significance of the UV range evident when drawing conclusions from the model fit in the UV and optical parts of the spectrum.

The parameter-fitting approach, on the other hand, simply postulates a velocity law (the only merit of which is that it shows qualitatively the right behavior), adapts the main parameter of that law (and an arbitrary mass loss rate) to obtain the best overall reproduction of the observed optical emission profiles, and then attributes everything it cannot fit to "clumping", without any additional constraints. How can anyone expect such an approach to *explain* anything about the physics of the atmosphere? What is its predictive power?

Even worse, the method runs the risk of attributing to clumping what is not caused by clumping, but rather by the difference between the density field assumed by the recipe and the one resulting from the actual hydrodynamics. It makes no sense to try to fit secondary effects such as clumping[5] as long as the primary influence (the shape of the underlying density and velocity fields) is not adequately considered. Still more critical is the overall choice of mass loss rate and terminal velocity. Because the essential connection between the stellar parameters and the wind properties (via the hydrodynamics of radiative driving) is missing in this approach, it simply does not allow any conclusions to be drawn about the stellar masses and luminosities from fits to the spectra, since for freely choosable mass loss rates and terminal velocities a plausible fit to an observed spectrum may be achieved even for completely wrong stellar masses and radii.

### 2.3. Further constraints

Further constraints on the nature of the driving mechanism of CSPN winds are offered by two other avenues of investigation. First, via a study of the behavior of the wind dynamics independent of the appearance of the spectra, and second, via modelling of the X-ray emission in the winds.

Regarding the first point it is important to keep in mind that the wind parameters are functions of the stellar parameters. Thus, if the wind parameters predicted for a set of stellar parameters contradict the observations, these stellar parameters must be wrong. Such an inconsistent behavior can, for example, be evident in the ratio $v_\infty/v_{\rm esc}$ of the terminal wind velocity and the escape velocity of individual stars. We have examined the $v_\infty/v_{\rm esc}$ ratios for our CSPN sample for two sets of stellar parameters – derived (a) from an analysis of the optical lines in combination with the post-AGB mass–luminosity relation (set 2), and (b) from analyses of the UV spectra together with consistent modeling of the winds (set 1) – and compared these to the corresponding ratios from a sample of massive O stars. The results showed that the spread in the $v_\infty/v_{\rm esc}$ ratios for our CSPN models of set 1 turned out to be comparable to that obtained from both massive O star observations and massive O star models, whereas the $v_\infty/v_{\rm esc}$ ratios of set 2 yielded much too high values incompatible with all other results (cf. Kaschinski et al. 2012). As the latter anomaly is coupled to the fact that the escape velocity $v_{\rm esc}$ is not a directly observable quantity (in contrast to $v_\infty$, which can be measured directly, $v_{\rm esc}$ depends on the stellar mass $M$ and radius $R$, and these quantities had been taken from the theoretical post-AGB mass–luminosity relation), we have to conclude from this result that there is a problem with the mass–luminosity relation at the CSPN high-mass end.

Although this is a striking result, it might possibly be construed as an intrinsic peculiarity of CSPN winds not exhibited by massive O star winds. It is therefore of great importance that the treatment of the expanding atmospheres of O-type stars offers another, independent, path of inquiry into the driving mechanism of CSPN winds with the modeling of the wind-produced X-rays. This is based on the hydrodynamical behavior of radiation-driven winds and can therefore be used as a further test of the reliability of the calculated wind dynamics.

As the X-rays are produced as a second-order effect by the cooling zones of shocked material within the winds, arising from the unstable, non-stationary behavior of the radiative wind driving mechanism (cf. Lucy & Solomon 1970, Owocki & Rybicki 1984), this kind of radiation can not only be observed directly, but also indirectly via its influence on the ionization balance. In particular, it leads to the production of higher ionization stages such as O VI (cf. Pauldrach 1987, Pauldrach et al. 1994b), which can be observed by means of the well-known O VI resonance doublet $\lambda\lambda$ 1031, 1037. To strengthen or weaken the general finding that CSPN winds are radiatively driven it is thus expedient to model the O VI line in the CSPN spectra and thereby to quantify the X-ray production rate and the radial distribution of the X-ray emitting gas, as has already been done for massive O star atmospheres (cf. Pauldrach et al. 1994a,b, 2001, 2012).

## 3. UV and optical data for the set of central stars of planetary nebulae

The UV observations redward of 1150 Å, acquired by the International Ultraviolet Explorer (IUE), were obtained from the IUE Final Archive data server (`http://sdc.laeff.inta.es/ines/`). Data for the spectral range blueward of 1150 Å, observed by the Far Ultraviolet Spectroscopic Explorer (FUSE), were obtained from the Multimission Archive (MAST) at STScI (`http://archive.stsci.edu/` – the data sets chosen for the CSPNs are: Tc 1: P198030400000; He 2-131: P193030100000; IC 418: P115111100000; IC 4593: D120030100000; NGC 2392: B032060100000; and NGC 6826: F160020900000). No postprocessing of the archival material was done apart from the standard procedure of rectification. The optical observations, acquired by a variety of telescopes and spectrographs (ESO 3.6 m + CASPEC, ESO NTT + EMMI, Isaac Newton 2.5 m (La Palma) + IDS, Palomar echelle, McDonald 2.1 m + Sandiford echelle), were kindly provided by R.-P. Kudritzki (priv. comm.).

## 4. An extended UV analysis of a selected sample of CSPNs

To investigate the influence of X-rays on the synthetic UV spectra of CSPNs and to analyze their characteristics we present in the following an extended UV analysis of a selected sample of CSPNs. With "extended" we mean that the UV analysis will not just be based on the comparison of our synthetic UV spectra with HST and IUE observations, but also with corresponding FUSE observations which include the wavelength range of the

[5] That the small inhomogeneities are indeed second-order effects has recently been shown by observations of small variations in the wind profiles of the CSPNs IC 418, IC 4593, and NGC 6826 (Prinja et al. 2012) – even small variations of the abundances produce stronger changes to the unsaturated wind line profiles (cf. Pauldrach et al. 2012) than the observed fluctuations. That the small amplitudes of the deviations from a smooth, stationary flow are second-order effects in general has also been discussed by Kudritzki (1999), and that the influence on the hydrodynamics of the outflow therefore is not marked has been verified by Pauldrach et al. (1994a) (see also references therein) and later on by Runacres & Owocki (2002). This behavior explains why the hydrodynamic models based on a smooth, time-averaged density structure are able to reproduce the multitude of UV spectral lines that are formed in the entire atmospheric depth range.

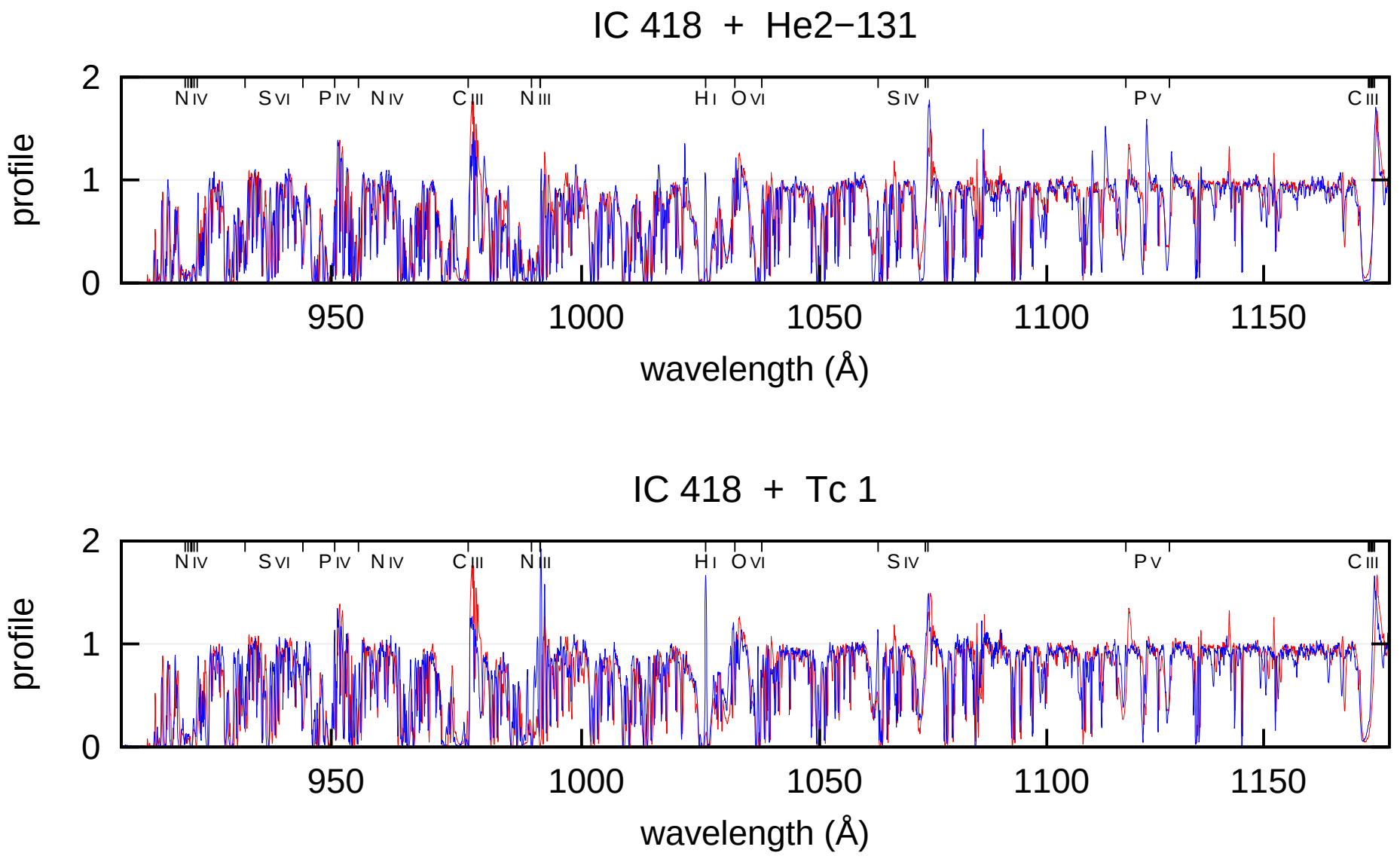

**Fig. 2.** Observed FUSE spectrum for IC 418 (red), compared to the observed spectra for He 2-131 (blue – upper panel) and Tc 1 (blue – lower panel). The comparisons show clearly that the behavior of the O VI line in the spectra of Tc 1 and He 2-131 is very similar to that of IC 418, and even the total FUSE spectra of these stars are not much different when compared to one another.

**Table 1.** Stellar masses and wind parameters for the six selected stars of our sample of central stars of planetary nebulae for which FUSE observations are available. The upper part lists the parameters derived from the consistent UV analysis (Pauldrach et al. 2004), the lower part the parameters obtained from the optical analysis (Kudritzki et al. 2006), the mass–luminosity relation, and consistently calculated dynamics.

| Object | $M$ ($M_\odot$) | $\dot{M}$ ($10^{-6} M_\odot$/yr) | $v_\infty$ (km/s) |
|---|---|---|---|
| Parameter Set 1 | | | |
| NGC 2392 | 0.41 | 0.018 | 400* |
| IC 4593 | 1.12 | 0.061 | 850 |
| IC 418 | 1.33 | 0.098 | 800 |
| (Tc 1 | 1.37 | 0.054 | 850) |
| NGC 6826 | 1.40 | 0.178 | 1200 |
| (He 2-131 | 1.39 | 0.321 | 400) |
| Parameter Set 2 | | | |
| NGC 2392 | 0.84 | 0.471 | 400* |
| IC 4593 | 0.70 | 0.120 | 250 |
| IC 418 | 0.92 | 0.639 | 200 |
| (Tc 1 | 0.84 | 0.605 | 300) |
| NGC 6826 | 0.74 | 0.187 | 600 |
| (He 2-131 | 0.71 | 0.105 | 300) |

* While Kudritzki et al. (1997), Pauldrach et al. (2004), Kudritzki et al. (2006), and Kaschinski et al. (2012) deduced a terminal velocity of 400 km/s for this object from the UV spectrum, Herald & Bianchi (2011), who also note that the absorption extends to 300 . . . 400 km/s, decided to use 300 km/s for the terminal velocity, interpreting the gradual absorption seen in some profiles to be the result of shocked material moving at higher velocities.

O VI resonance line. This spectral line is primarily produced by Auger ionization driven by X-rays emitted by the shock cooling zones resulting from the unstable, non-stationary behavior of the winds, and thus represents a special diagnostic tool for these shocks by allowing their consequences to be analyzed from observations in the UV spectral range, in particular regarding the quantity and the structure of the X-rays (Pauldrach et al. 1994b, 2001, 2012). With the FUSE data available this diagnostic potential of the O VI line can be utilized for six CSPNs of our sample: NGC 2392, IC 4593, IC 418, Tc 1, NGC 6826, and He 2-131.

For these six CSPNs we list in Table 1 the masses and wind parameters (all other stellar parameters of these objects are given in Table 3) obtained from the different analysis techniques (cf. Sect. 2): in the upper part the parameters deduced from the UV analyses (parameter set 1; Pauldrach et al. 2004) and in the lower part the parameters obtained from the optical analyses (Kudritzki et al. 2006), the mass–luminosity relation, and a consistent treatment of the hydrodynamics (parameter set 2). Examples of comparisons of the UV spectra from WM-basic model runs to the observations are shown in Figs. 3 through 6 (for the four CSPNs NGC 2392, IC 4593, IC 418, and NGC 6826; the behavior of the O VI line in the spectra of Tc 1 and He 2-131 (not shown) is very similar to that of IC 418 – cf. Fig. 2).

**The CSPN NGC 2392.** The middle panel of Fig. 3 shows the synthetic UV spectrum of our standard model (parameter set 1) of NGC 2392. The predicted spectrum is in good agreement with not only the observed IUE spectrum, but also with the FUSE observations. (The small deviations seen in a few spectral lines can of course be corrected by simply adjusting the corresponding abundances slightly. As we discuss below, such fine-tuning is beyond the scope of our present investigation.) The observed O VI line, however, is clearly not reproduced by the synthetic spectrum, and this failure indicates strongly that – as in massive O stars – X-rays are also an important ingredient in CSPN winds. As X-rays are produced by shock cooling zones embedded in the wind, the physics that describe this phenomenon must thus be applied to CSPN models in the same manner as it has been applied to the models of massive O stars. (The primary effect of the EUV and X-ray radiation is its influence on the ionization equilibrium, where the enhanced direct photoionization due to the EUV shock radiation is as important as the effects of Auger ionization caused by the soft X-ray radiation (cf. Pauldrach 1987, Pauldrach et al. 1994b). The radiation is explained by radiative losses of post-shock regions from shocks pushed by non-stationary features, and our modeling is based on a stationary "cool wind" with embedded randomly distributed shocks where the shock distance is much larger than the shock cooling length in the accelerating part of the wind and only a small amount of high-velocity material appears, with a filling factor usually smaller than $f \approx 10^{-2}$ and jump velocities of about $u_s \approx 200 \ldots 800$ km/s which result in immediate post-shock

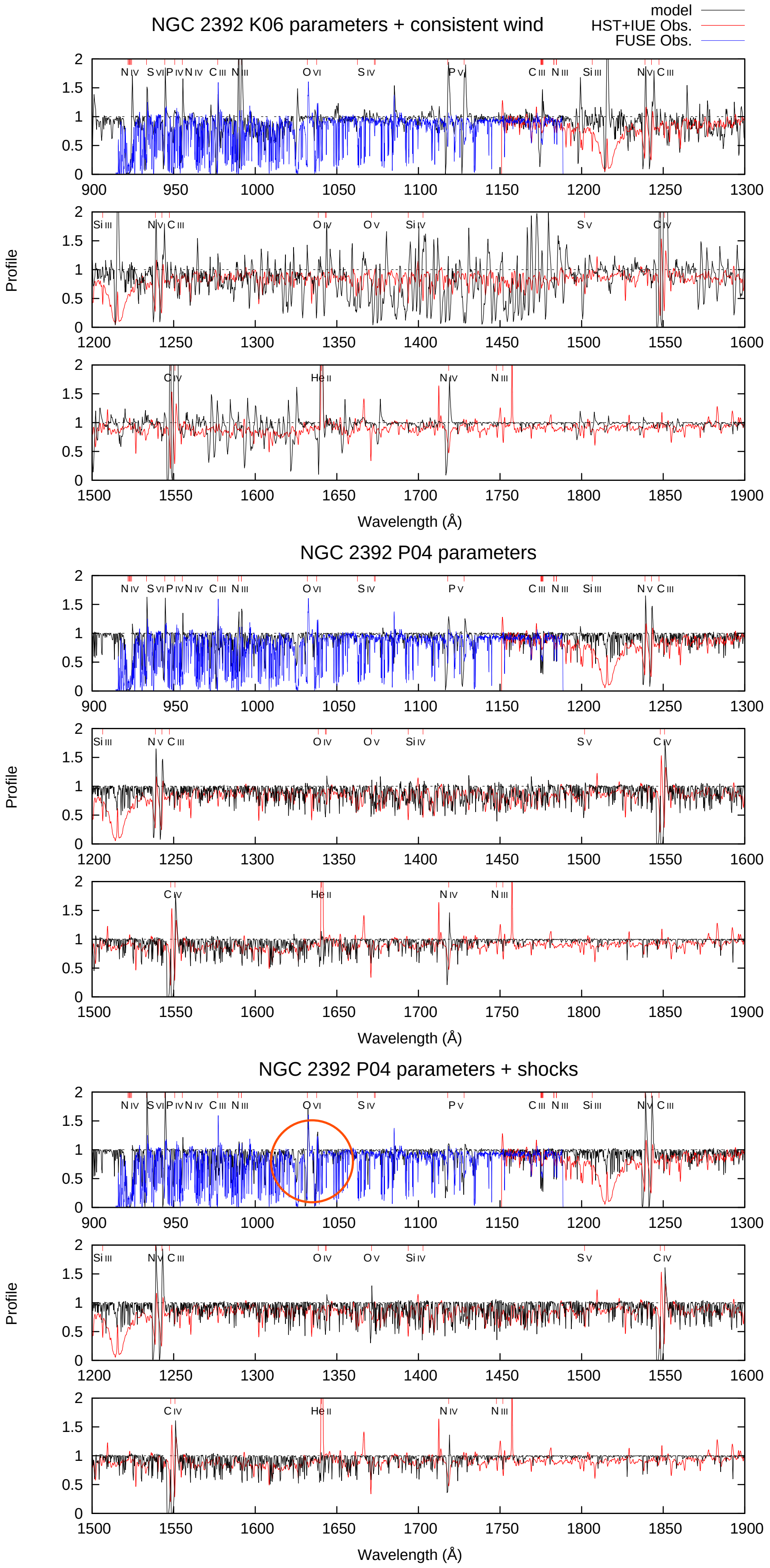

**Fig. 3.** Synthetic UV spectra (black) from consistent models for NGC 2392, compared to the observed UV spectra (red and blue). The top panel show the spectrum from a model using the stellar parameters from the analysis of Kudritzki et al. (2006). The spectrum is clearly incompatible with the observations, thus ruling out this parameter set. The center panel shows the spectrum from a model using the parameters from the analysis of Pauldrach et al. (2004). The spectrum predicted by the model is generally in good agreement with the observations, but the O VI line is not reproduced. In the bottom panel, shocks in the wind have additionally been included in the modelling procedure. The small amount of high-energy radiation ($L_X/L_{bol} \sim 10^{-6}$) generated by these shocks is enough to ionize a small fraction of oxygen into O VI, sufficient to reproduce the observed line strength. This result indicates strongly that shocks and the X-ray radiation emitted by them play a role not only in the winds of massive O stars, but also in the winds of CSPNs.

**Table 2.** Shock structures of the four central stars: NGC 2392, IC 4593, IC 418, and NGC 6826. These have been determined from fits of the calculated O VI lines to the FUSE observations. The table lists characteristic values of the radial distributions of the maximum local shock temperatures $T^{\max}_{\rm shock}(r)$ and the integrated relative X-ray luminosities $L_X/L_{\rm bol}$ emitted by the shock cooling zones embedded in the winds.

| | | | | | | | |
|---|---|---|---|---|---|---|---|
| NGC 2392 | | IC 4593 | | IC 418 | | NGC 6826 | |
| $\log(L_X/L_{\rm bol}) = -6.0$* | | $\log(L_X/L_{\rm bol}) = -6.3$* | | $\log(L_X/L_{\rm bol}) = -6.6$ | | $\log(L_X/L_{\rm bol}) = -6.4$ | |
| r ($R_*$) | $T^{\max}_{\rm shock}(r)$ ($10^6$ K) | r ($R_*$) | $T^{\max}_{\rm shock}(r)$ ($10^6$ K) | r ($R_*$) | $T^{\max}_{\rm shock}(r)$ ($10^6$ K) | r ($R_*$) | $T^{\max}_{\rm shock}(r)$ ($10^6$ K) |
| 1.202 | 0.255 | 1.110 | 0.054 | 1.118 | 0.040 | 1.094 | 0.045 |
| 1.309 | 0.347 | 1.226 | 0.117 | 1.318 | 0.073 | 1.416 | 0.143 |
| 1.529 | 0.495 | 1.381 | 0.227 | 1.573 | 0.131 | 1.807 | 0.289 |
| 1.748 | 0.606 | 1.746 | 0.507 | 1.829 | 0.188 | 2.197 | 0.415 |
| 2.182 | 0.760 | 2.112 | 0.755 | 2.381 | 0.291 | 3.421 | 0.679 |
| 2.669 | 0.873 | 4.184 | 1.546 | 3.067 | 0.384 | 5.754 | 0.915 |
| 4.287 | 1.063 | 5.393 | 1.768 | 4.728 | 0.515 | 9.344 | 1.063 |
| 8.845 | 1.226 | 11.31 | 2.206 | 9.265 | 0.651 | 18.12 | 1.187 |
| 27.08 | 1.328 | 29.05 | 2.472 | 27.41 | 0.755 | 41.51 | 1.264 |
| 51.38 | 1.352 | 52.70 | 2.551 | 51.61 | 0.781 | 53.21 | 1.278 |
| 75.69 | 1.360 | 76.35 | 2.581 | 75.80 | 0.791 | 76.61 | 1.292 |
| 100.0 | 1.365 | 100.0 | 2.597 | 100.0 | 0.796 | 100.0 | 1.300 |

* We note that based on a less elaborated treatment of shock structures Herald & Bianchi (2011) obtained integrated relative X-ray luminosities $L_X/L_{\rm bol}$ which for the central stars NGC 2392 ($L_X/L_{\rm bol} = -4.4$), and IC 4593 ($L_X/L_{\rm bol} = -4.8$) are almost a factor of 100 higher than our values. (Their parameters have also been determined from fits of the calculated O VI lines to the observations). However, as pointed out by the authors themselves, their results will certainly be revised by the application of a more sophisticated treatment of X-rays, which they state is an ongoing development.

temperatures of approximately $T_s \approx 10^6 \ldots 10^7$ K – cf. Pauldrach et al. 1994a.[6])

In the lower panel of Fig. 3 we show the synthetic spectrum from our model of NGC 2392 in which we have included shocks in this manner. The spectrum now reproduces the profile of the O VI P-Cygni line quite well, and as an integral part of the modelling procedure we obtain the emitted frequency-integrated relative X-ray luminosities[7] $L_X/L_{\rm bol}$ as well as the radial distribution of the maximum local shock temperatures[8] $T^{\max}_{\rm shock}(r)$ in the cooling zones of the shock-heated matter component (Table 2).

While the synthetic spectrum of our current best model (whose parameters had been determined from an analysis of these spectra) is now quite well in agreement with the complete observed UV spectrum – the IUE and HST observations along with the FUSE data –, the same cannot be said of the model shown in the upper panel, which does not at all reproduce the observations. This model has been calculated using parameter set 2 (obtained from the optical analyses and the theoretical post-AGB mass–luminosity relation) together with consistently calculated dynamics. The failure to reproduce the observations is mainly due to the much too high mass loss rate resulting from these stellar parameters, although the resulting terminal velocity is close to the observed value. Because spectra computed with such stellar parameter sets – obtained from optical analyses and corresponding consistent wind parameters – showed large discrepancies when compared to observed spectra (in the optical as well as in the UV regimes) in the past, Kaschinski et al. (2012) had already concluded that *the published optical analyses give good fits to the observed spectrum only because the wind parameters assumed in these analyses are inconsistent to their stellar parameters*. With regard to the comparison shown in the upper panel of Fig. 3 we reach the same conclusion here (see also the upper panels of Figs. 4, 5, and 6).

**General considerations regarding the spectral diagnostics.** As can be seen when comparing the UV spectra of the two models shown in the upper and lower panel of Fig. 3, there are no particular selected features which react to the mass loss rate, but

[6] On this basis we account for the density and temperature stratification in the shock cooling layer and the two limiting cases of radiative and adiabatic cooling layers behind shock fronts by replacing the usual volume emission coefficient $\Lambda_\nu(T_s(r), n_e)$ by integrals over the cooling zones denoted by $\hat\Lambda_\nu(T_s(r))$. For the shock emission coefficient $\epsilon^s_\nu(r)$ we thus get

$$\epsilon^s_\nu(r) = \frac{f}{4\pi} n_p n_e \hat\Lambda_\nu(T_s(r)),$$

where

$$\hat\Lambda_\nu(T_s(r)) = \pm\frac{1}{x_s}\int_r^{r\pm x_s} \hat f^2(r')\Lambda_\nu(T_s(r')\,\hat g(r'))\,dr',$$

$n_p n_e$ is the product of the number densities of the ions and electrons, and $r$ is the radial location of the shock front. The plus sign corresponds to forward and the minus sign to reverse shocks, $r'$ is the cooling length coordinate with a maximum value of $x_s$, and $\hat f(r')$ and $\hat g(r')$ denote the normalized density and temperature structures with respect to the shock front (cf. Pauldrach et al. 2001).

[7] As is customary, we list the frequency-integrated X-ray luminosity $L_X$ normalized to the bolometric luminosity $L_{\rm bol}$; this gives the fraction of the total luminosity emitted in the X-ray regime. Note that most of the locally produced X-ray radiation is reabsorbed by the wind, but a small fraction escapes and is observable as soft X-rays with $L_X/L_{\rm bol} \approx 10^{-7}$. Apart from the predicted influence on lines formed within the wind, the accuracy of the shock description can thus be additionally verified by a comparison to direct X-ray observations.

[8] The maximum local shock temperatures $T^{\max}_{\rm shock}(r)$ are given by the Rankine-Hugoniot relation as function of the immediate jump velocities $v_{\rm shock}(r)$. Normalized to the turbulent velocity $v_{\rm turb}$, which can be determined from the fit of the shape of the P-Cygni line profiles, the local values of the jump velocity $v_{\rm shock}(r)$ can be estimated via a correlation of $v_{\rm shock}(r)/v_{\rm turb}$ to the wind outflow velocity $v(r)/v_\infty$ (Pauldrach et al. 1994b).

rather practically *all* of the lines in the UV change with varying stellar parameters – a wavelength interval of, say, 500 Å in the UV range covers hundreds of strong and equally important lines which the consistent UV analysis tries to fit simultaneously.

The primary aim of the current paper is to judge proposed sets of stellar parameters by their influence on the appearance of the spectra. As the comparisons between the upper and the lower panels of Figs. 3, 4, 5 and 6 show, the spectra change considerably from one set of stellar parameters to the other, and one of the two *clearly* results in a *much* better fit than the other, even though that fit is perhaps not perfect in all details. What matters at this point, however, is the huge possible range over which the spectrum can actually vary. *Thus, for the purposes of the present analysis to judge the stellar parameters, it is sufficient to present the computed spectra on a scale where the response of the spectrum as a result of varying the input parameters becomes visible.*

We stress that our work here represents the first step of an analysis, where the basic parameters including the mass and the luminosity of an object are determined via a consideration of the fundamental physics. This is different from a second step, where through fine-tuning of the abundances more information is attempted to obtain from the line fits (this regards especially the S VI, P V, N V, C IV, and He II line). Although it is tempting to present the second step before – or instead of – the first one (since obtaining results that look impressive simply by fixing free parameters is comparatively effortless), performing this second step is beyond the objective of the present paper.

**The CSPNs IC 4593, IC 418, and NGC 6826.** The findings and conclusions we have obtained from the UV analysis of NGC 2392 are also reflected in the UV analyses of the central stars IC 4593, IC 418, and NGC 6826. As is shown in the lower panels of Figs. 4, 5, and 6 the O VI lines of these stars are, along with the rest of the spectra, quite well reproduced by our method and especially by our treatment of the shock physics (see also Fig. 1). (The parameters which describe the behavior of the shocks and which are deduced from the analysis of the O VI line are shown in Table 2). In contrast to that, the upper panels of Figs. 4, 5, and 6 show fits of the UV spectra of these CSPNs that are of poorer quality, reflecting the weakness of the used model parameter sets 2 to represent these objects realistically. As in the case of NGC 2392 the primary reason for this behavior are the consistently determined mass loss rates which give too high values for all of these objects. But in contrast to the model of NGC 2392 the obtained terminal velocities are also too small for these three CSPNs. (With respect to these striking mismatches it does not make much sense to also apply the treatment of shocks to these models). This result implies an additional strong hint that something is wrong with these parameter sets (sets 2 in Table 1) – and this certainly regards the stellar masses which have not been determined spectroscopically but assumed in this case in order to complete the sets of stellar parameters.

**Discussion of the characteristic values of the X-ray emission.** As a consequence of our investigation, which revealed the behavior of the X-ray emission of our sample of CSPNs and thus the structure of the underlying shock physics constrained by the quality of the fits of the O VI line, the deduced values of the integrated X-ray luminosities and the radial distribution of the maximum local shock temperatures can now also be compared with corresponding values resulting directly from observations. Kastner et al. (2012) have presented results obtained from the first systematic Chandra X-ray Observatory survey of planetary nebulae (PNe) in the solar neighborhood (within ∼1.5 kpc). The highlights of these results include the finding that roughly 50% of the PNe observed by Chandra harbor X-ray-luminous CSPNs, and that this fraction involves detections of diffuse X-ray emission and detections of X-ray point sources[9]. Most noticeable is that five objects, including NGC 2392, display both diffuse and point-like emission components in the Chandra imaging. From an observational point of view it is therefore imperative to disentangle the origin of the hot gas, since it consists of two components: the shocked fast stellar wind (diffuse source) and the emission from its central star (point source).

In this regard it is quite interesting that Guerrero et al. (2005) derived for NGC 2392 and its surrounding Eskimo Nebula at our distance of 1670 pc an X-ray luminosity of $L_\mathrm{X} = (2.6 \pm 1.0) \times 10^{31} \cdot (1670/1150)^2 = 5.4 \times 10^{31}$ erg s$^{-1}$ which is almost exactly a factor of two larger than the value $L_\mathrm{X} = 2.3 \times 10^{31}$ erg s$^{-1}$ we have deduced from our best-fit model for this CSPN (cf. Table 1 and Table 2). This result is not only in accordance with the observational fact of Kastner et al. (2012) (cf. their Fig. 3), but also with the interpretation of Guerrero et al. (2005) of their result: They noted that the X-ray emission of NGC 2392 has to be partially attributed to the central star and partially to the diffuse X-ray emission produced by the dynamical interaction of the fast outflow with the inner shell of the PNe. As Chandra has detected a relatively hard point source emission in NGC 2392 (Kastner et al. 2012), which likely explains the energy dependence of the X-ray morphology apparent in the earlier XMM data (cf. Guerrero et al. 2005), the significance of this encouraging result can certainly be improved by a comparison of the maximum shock temperatures. As shown, the best-fit model of Kastner et al., overplotted on the EPIC/pn spectrum in their Fig. 2, has a plasma temperature of around $2.0 \times 10^6$ K, which compared to our value of $1.4 \times 10^6$ K (cf. Table 1) is somewhat larger, but the highest values of the temperature of the shocked gas is definitely due to the highest Mach-number to be expected and this has to be found in the diffuse component and not the point source. We therefore regard our maximum shock temperature to be also in accord with the current observations.

Our general finding that the X-ray to bolometric luminosity ratio for CSPNs is in the range $L_\mathrm{X}/L_\mathrm{bol} \sim 10^{-7} \ldots 10^{-6}$ (cf. Table 1) is further confirmed by Guerrero (2006), who deduced for the central star NGC 6543 a value of $L_\mathrm{X}/L_\mathrm{bol} \sim 10^{-7}$ from observations. As such values of $L_\mathrm{X}/L_\mathrm{bol}$ are typical for radiation driven winds – they are of the same order as observed for massive O stars (cf. Sana et al. 2005 and Sana et al. 2006) –, it is obviously the radiation pressure which determines the atmospheric structure not just of massive O stars but also of O-type CSPNs; and this regards not only the primary behavior which rules the values of the mass loss rate $\dot{M}$ and the terminal velocity $v_\infty$, but also weak and secondary effects which rule the details of the shock physics.

Our consistent procedure and the corresponding analysis is thus not only based on the high radiation flux density of CSPNs (which necessarily leads to expanding atmospheres), the wind-momentum–luminosity relation (WMLR), and the behavior of $v_\infty/v_\mathrm{esc}$, which can be compared to observations (cf. Pauldrach et al. 2004 and Kaschinski et al. 2012), but also on secondary effects such as the maximum shock temperatures and the integrated X-ray luminosities which are also subject of observations.

[9] The diffuse X-ray emission is thought to arise from shells where the CSPN wind rams into the previously expelled AGB envelope, producing hot shocked gas; the point-like sources are associated with the central stars themselves.

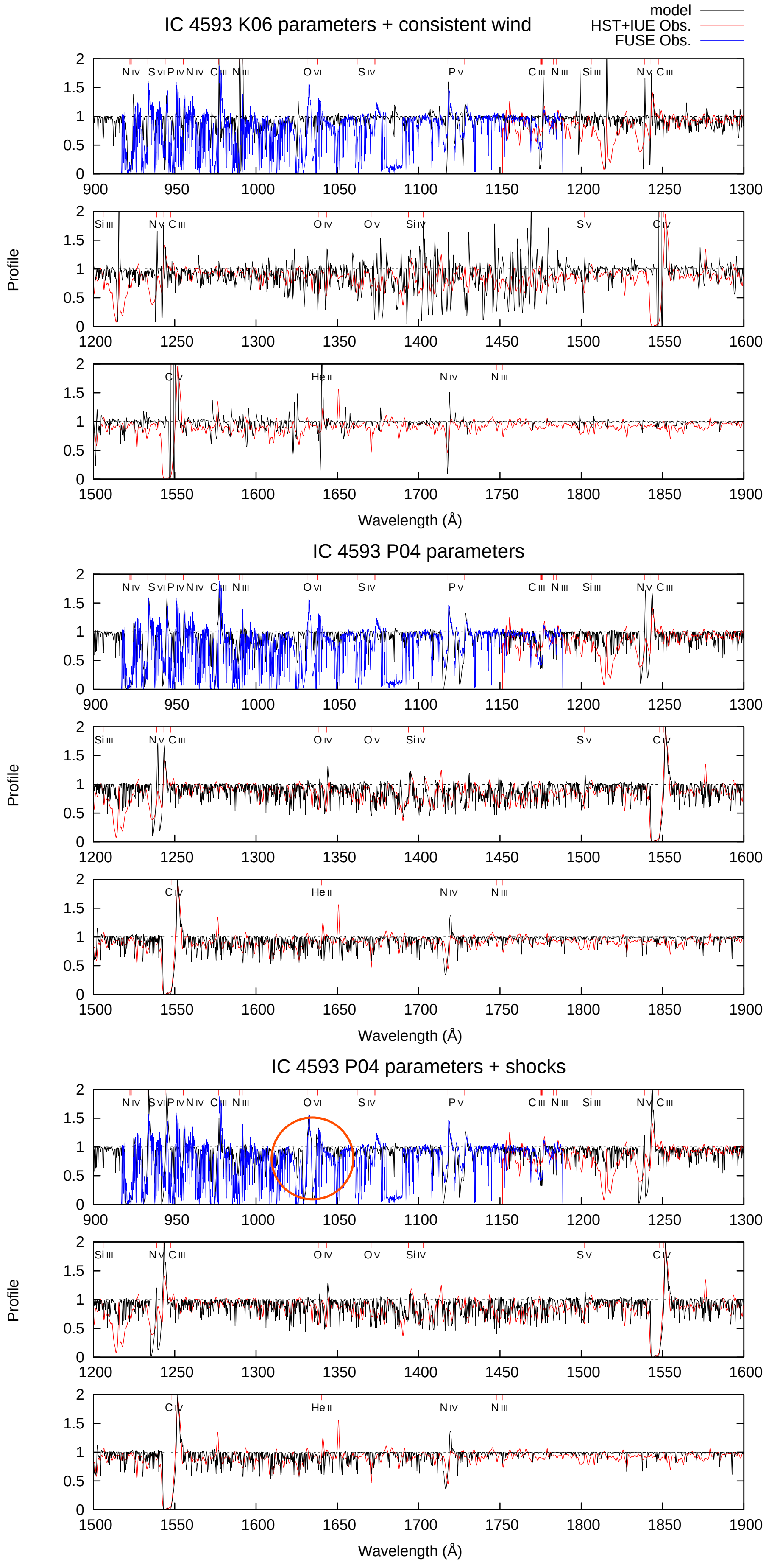

**Fig. 4.** As Fig. 3, but for IC 4593.

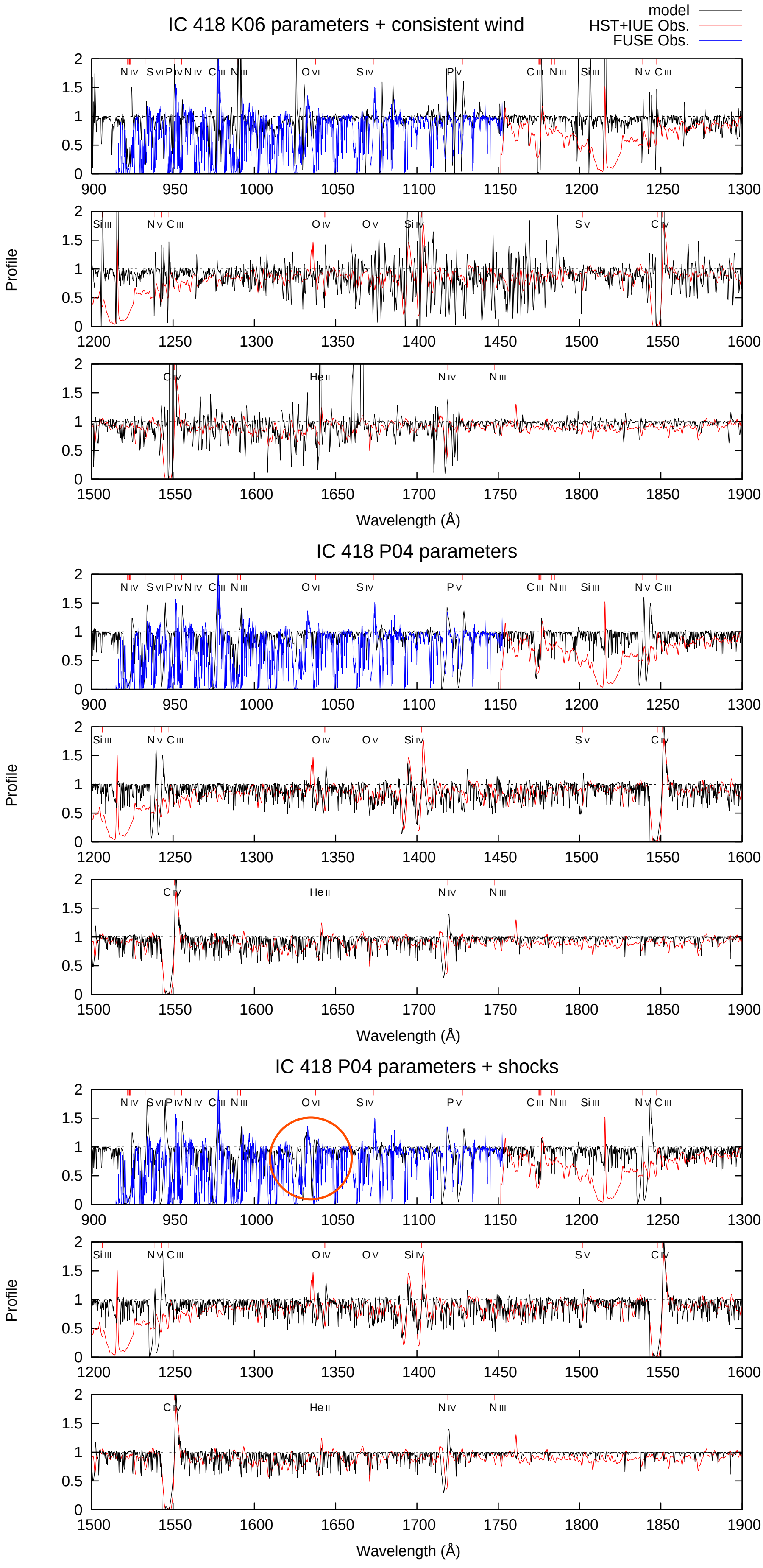

**Fig. 5.** As Fig. 3, but for IC 418.

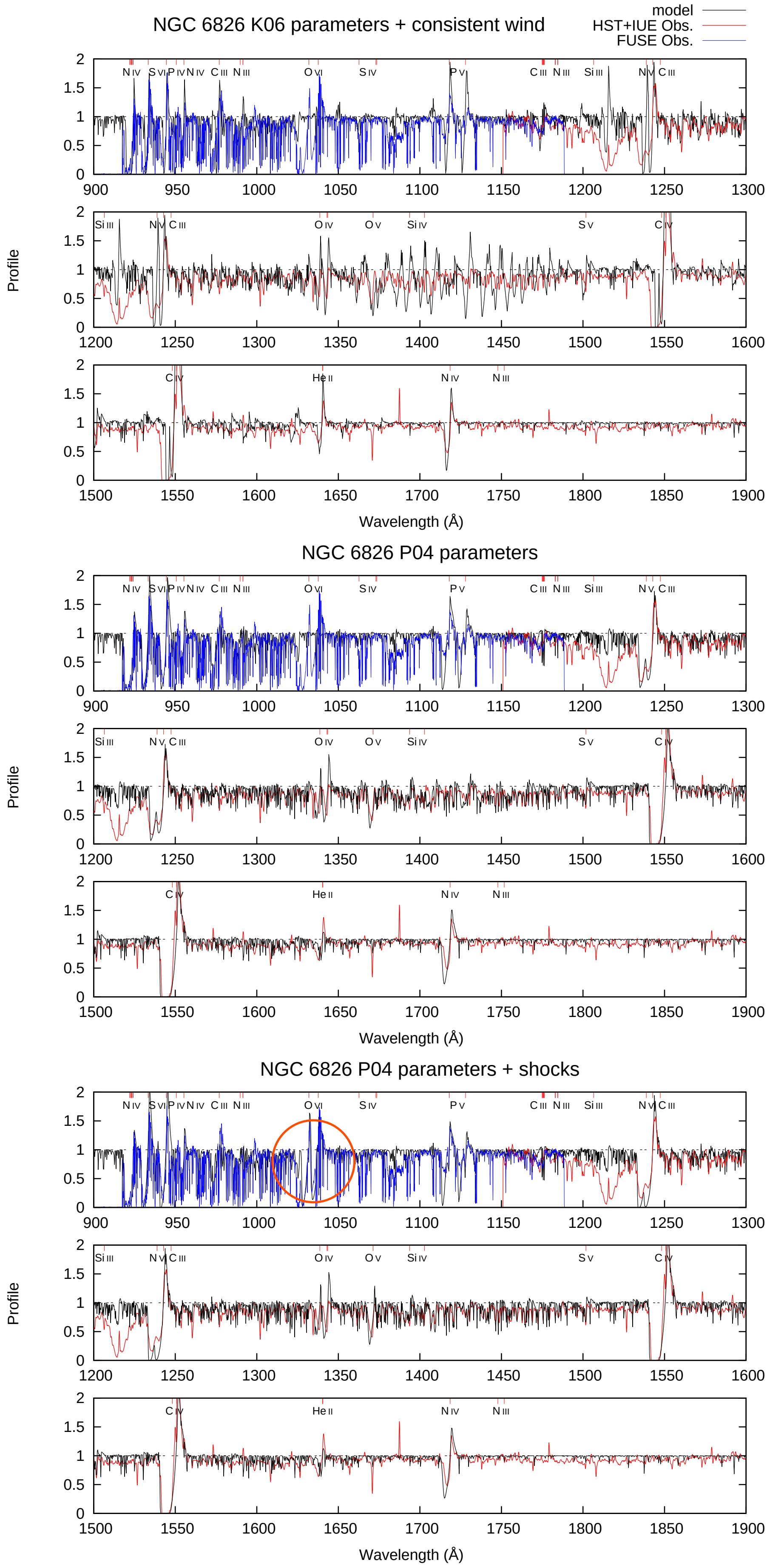

**Fig. 6.** As Fig. 3, but for NGC 6826.

All these findings support the hypothesis that the atmospheres of CSPNs are governed by exactly the same radiative line-driving mechanism as those of massive O stars.

## 5. Accounting for clumping (?)

Although the spectrum in the optical wavelength range is also computed by our code, it is not our means of spectral diagnostics, because compared to the UV spectral range it contains only few and only rather weak subordinate lines (cf. Fig. 1) which are not only much more susceptible to uncertainties in the modelling than the lines in the UV, but moreover may be strongly influenced (in particular H$\alpha$ and He II $\lambda$ 4686) by second-order effects possibly connected to small inhomogeneities in the density. Such inhomogeneities, leading to stronger emission from H and He recombination lines due to the higher densities in the "clumps" compared to a smooth outflow, may be the reason why Kaschinski et al. (2012) could not fit the H$\alpha$ and the He II $\lambda$4686 lines as well as the rest of the spectrum.

Inhomogeneities, in fact, are the currently favored explanation for such discrepancies. They are often described using ad-hoc "clumping factors" employed in the models to bring the predicted H$\alpha$ and He II $\lambda$4686 lines into agreement with the observed profiles. There is, however, a certain caveat attached to this treatment. Supposing that clumping (or some other effect with similar influence on these two lines) does indeed play a role, then – as long as we have no consistent physical description for this effect – we have complete freedom to fit the optical emission lines without gaining any additional information about the stellar parameters.[10] But a line whose model profile is determined primarily by arbitrarily adapting clumping factors and not the underlying physics of the model completely loses its diagnostic value, and the quality of the fit of this line says nothing about the reliability of the fundamental model parameters.[11] Such a procedure furthermore involves the risk of covering up intrinsic weaknesses in other parts of the model description.

A consistent description of clumping, which would really mark progress in the field, has not yet been developed, however. Moreover, most work to date on clumping has unfortunately been based on similar ad-hoc parametrizations for the velocity field, thus possibly confusing the issue further. Although the first attempts to consistently compute the velocity field date back to at least 1975 (Castor, Abbott, & Klein), since then there has been very little work published worldwide in this area. This is astonishing, since a further development of the field to include time dependence and deviations from spherical symmetry would promise much more fundamental insights than ad-hoc parametrizations, and through this ultimately lead to a consistent theory of clumping. (As the usage of ad-hoc methods for the clumping factors has become an accepted practice over the last 10 years, it would almost appear as if most researchers in the field shunned the really ambitious and challenging subjects, and instead content themselves with playing around with second-order effects that require much less effort.)

We must therefore caution against reading too much into the deviations of these few lines, and at the present stage it is highly questionable whether the affected hydrogen and helium lines should be regarded as meaningful observables from which reliable information about the stellar and wind parameters can be deduced. On the other hand, a consistent theory of clumping is only required if the affected spectral lines are actually used diagnostically to constrain the basic model parameters. For our present purposes this is not the case: the models are constrained by the physics of radiative driving and the appearance of the UV spectra. Thus, on the basis of the primary influence due to the (independently predicted) velocity field, these lines might therefore be possibly used to quantify the influence of small inhomogeneities on the shapes of the concerned lines. We investigate in the following whether wind clumping may constitute a plausible explanation for the fact that the models of Kaschinski et al. (2012) based on a hydrodynamically consistent smooth outflow could not reproduce the strength of the optical emission lines H$\alpha$ and He II $\lambda$ 4686. We investigate further to which degree this would influence the UV spectra on which the parameter determinations had been based. For this investigation we will use a simple ad-hoc description of clumping similar to that advocated by others.

To simulate the effects of clumping we consider the same basic treatment as implemented in FASTWIND (Puls et al. 2006), the model atmosphere code that had been used by Kudritzki et al. (2006) in their analysis. In this "microclumping" description the gas is considered to be swept up into small, optically thin clumps with a volume filling factor of $1/f_{\mathrm{cl}}(r)$ and void interclump regions. The density in the clumps,

$$\rho_{\mathrm{cl}}(r) = f_{\mathrm{cl}}(r)\bar{\rho}(r), \tag{1}$$

is then the local clumping factor $f_{\mathrm{cl}}(r)$ times the density $\bar{\rho}(r)$ of the smooth, unclumped flow given by the equation of continuity,

$$\bar{\rho}(r) = \frac{\dot{M}}{4\pi r^2 v(r)}. \tag{2}$$

The void interclump regions do not influence the radiative intensity, and only the fraction $1/f_{\mathrm{cl}}$ of volume that actually contains matter contributes to the opacity and emissivity along a macroscopic path element. Thus, if $\kappa = \chi/\rho$ denotes the opacity per unit mass, the optical depths

$$\tau = \int \chi_{\mathrm{cl}} \frac{1}{f_{\mathrm{cl}}}\,\mathrm{d}s = \int \kappa\rho_{\mathrm{cl}} \frac{\mathrm{d}s}{f_{\mathrm{cl}}} = \int \kappa f_{\mathrm{cl}}\bar{\rho} \frac{\mathrm{d}s}{f_{\mathrm{cl}}} = \int \kappa\bar{\rho}\,\mathrm{d}s = \int \bar{\chi}\,\mathrm{d}s, \tag{3}$$

and indeed all quantities that are linear in the density, remain unaffected by this clumping except that clumping may lead to a shift in the ionization stages, and thus a change in $\kappa$, through increased recombination in the higher-density clumps. On the other hand, quantities that are proportional to the square of the density, such as the emissivities of recombination lines, are enhanced by a factor of $f_{\mathrm{cl}}$ compared to an unclumped flow with the same mean density.

To demonstrate that our clumping treatment in WM-basic produces results comparable to FASTWIND, we show in Fig. 7 the predicted H$\alpha$ line profiles[12] for two stellar models with strong winds (in which H$\alpha$ is most likely to be affected by clumping). We use, here as well as in the CSPN model calculations of Sect. 5.1, the same radial run of the clumping factors as used by Kudritzki et al. (2006) in their analysis (J. Puls, priv. comm.): no clumping (i.e., $f_{\rm cl}(r) = 1$) in the photosphere and up to $v = 0.05 v_\infty$; a steep rise to the specified clumping factor from $v = 0.05 v_\infty$ to $v = 0.1 v_\infty$; and then keeping this value $f_{\rm cl}(r)$ = const for $v > 0.1 v_\infty$. To focus the comparison on the clumping and the radiative transfer we have set up the comparison runs as closely as possible, in particular we have used the density structures of the FASTWIND models in our WM-basic calculations. The remaining small differences in the predicted line profiles are due to actual differences in the modelling, such as the use of different model atoms, or the different interpolation of the opacities and emissivities between the grid points in the radiative transfer.

[10] Note also that without a description of clumping based on first principles, clumping is not a single fit factor but a whole number of them, one for each depth point of the model, and thus the degree of freedom is extremely large. Using these parameters it might be possible to fit the observed optical emission lines even with a wrong underlying velocity and density field.

[11] Although the names "clumping" and "porosity" suggest (correctly) that they are based on real physical ideas, with the presently missing quantitative connection to the other physical parameters of the model (such as radiative flux and line-driving acceleration) they are currently just other words for "free parameters to make the spectrum look nice".

[12] We had already shown in Kaschinski et al. (2012) that the photospheric absorption lines, which are formed deep in the atmosphere where the density approaches a hydrostatic structure and is therefore independent of the density law used to describe the wind in the outer

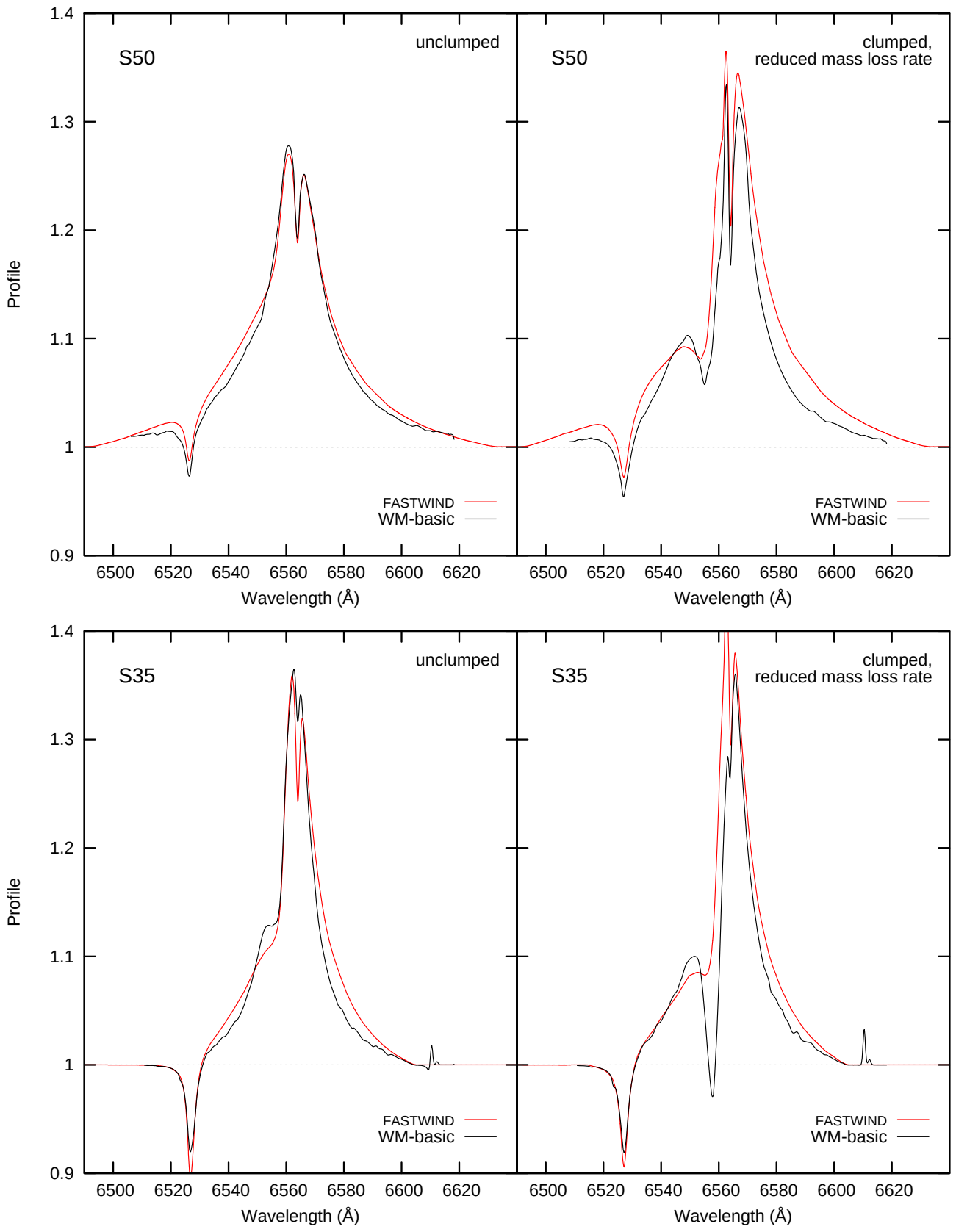

**Fig. 7.** Comparison of H$\alpha$ line profiles predicted by FASTWIND (red) and WM-basic (black). To focus the comparison on the NLTE modelling and the treatment of radiative transfer, these WM-basic calculations use the density structures from the FASTWIND models. The top row shows a $T_{\rm eff}$ = 50 000 K test model (S50), the bottom row a $T_{\rm eff}$ = 35 000 K test model (S35). The left panels show the profiles from a smooth, unclumped flow ($f_{\rm cl}$ = 1), the right panels those from a clumped flow ($f_{\rm cl}$ = 81) with a correspondingly reduced mass loss rate (by a factor of $\sqrt{f_{\rm cl}}$ = 9). The differences between our WM-basic predictions and those from FASTWIND are negligible compared to the sensitive dependence of the H$\alpha$ line on the clumping factor, as illustrated by Fig. 8.

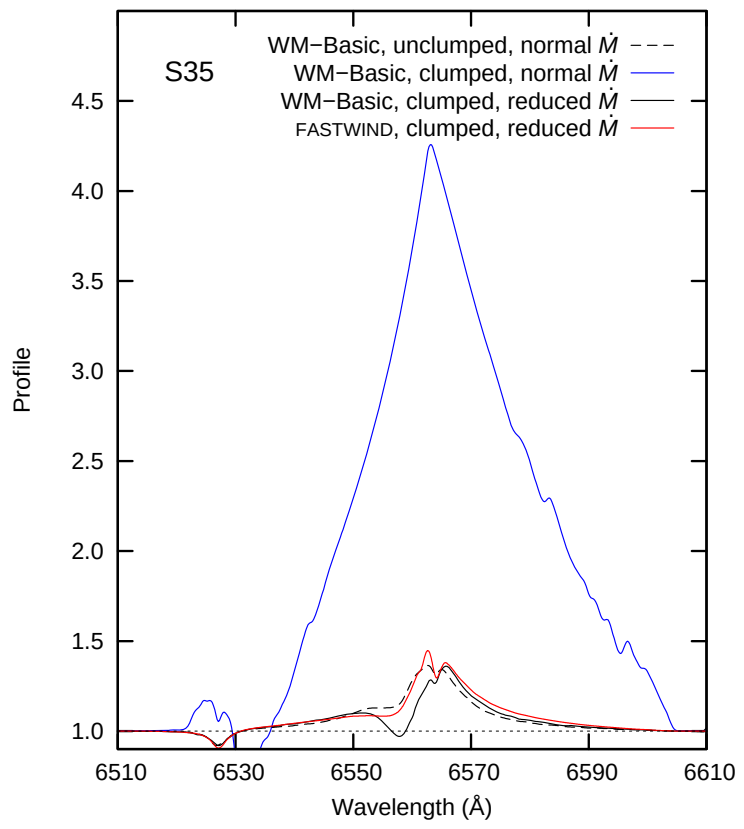

**Fig. 8.** H$\alpha$ line profile of the S35 test model as in Fig. 7, but this time also showing the case of a clumped wind ($f_{\rm cl}$ = 81, as in Fig. 7) with the "normal" mass loss rate (blue). Compared to this strong response, the small differences between two corresponding FASTWIND (red) and WM-basic (black solid line) predictions are insignificant.

These small differences in the predicted line profiles are not relevant, however, for the purposes of diagnosing clumping factors from given observed profiles. This is demonstrated in Fig. 8 where we compare the predicted H$\alpha$ line profiles from an unclumped wind and from a clumped wind ($f_{\rm cl}$ = 81) with the same mass loss rate. (The latter is equivalent to an unclumped wind with a mass loss rate increased by a factor of $\sqrt{f_{\rm cl}}$.) Compared to the strong response of the profile to a change in the mass loss rate or clumping factor, the differences in the predictions of WM-basic and FASTWIND are small. In other words, for a given clumping factor one would deduce exactly the same mass loss rate from an observed H$\alpha$ profile when using the WM-basic model as when using the FASTWIND model.

atmospheric regions, are very similar between FASTWIND and WM-basic. We therefore refrain from showing those profiles here again and instead remark only that they are also very similar for the benchmark models presented here.

In these tests (and these tests only) we have used the density structures from the FASTWIND models in our WM-basic calculations. We put particular emphasis on this point, since we have shown here that both codes produce equivalent line profiles *if given the same input*. This is important to keep in mind for the model calculations for individual CSPNs below: when using the stellar parameters of Kudritzki et al. (2006), if we do not achieve as good a fit to the observed profiles as they did, *even when enforcing their mass loss rates and terminal velocities* in our models, then this is not because the codes a priori produce different results, but because a *different density structure* was used. *But that is the entire point of using our modelling code in these analyses: to base the model predictions on a density structure that follows from a solution to the equation of motion of the wind, instead of simply* assuming *a certain parametrization*[13] as Kudritzki et al. (2006) did. For the H$\alpha$ line profile formation the differences between the assumed and the computed wind density structures play at least as important a role as the proposed clumping.

### 5.1. Model calculations for individual CSPNs

In the following we present a comparison of observed and computed optical line profiles for our sample of nine CSPNs. Basis for this comparison are the two sets of stellar parameters derived (a) from an analysis of the UV spectra by Pauldrach et al. (2004), hereafter labeled P04, and (b) from an analysis of the optical spectra by Kudritzki et al. (2006), labeled K06. The parameters of all objects in the sample are listed in Table 3.

[13] If we *knew* that the winds were completely smooth, a point could be made for "empirically" determining the velocity law by trying different parameter values in the prescription and choosing the one that leads to the best fit of the H$\alpha$ line profile, but once we admit clumping, for which we have no theory to predict the radial run either, such reasoning becomes rather specious.

**Table 3.** Overview of models for the nine CSPNs of the investigated sample. In the parameter source column, P04 refers to the parameters derived by Pauldrach et al. (2004) from an analysis of the UV spectra, and K06 refers to the optical analysis of Kudritzki et al. 2006. "Consistent" means that the wind parameters are consistent with the stellar parameters as determined by our hydrodynamic calculations of the radiative driving force.

| Object | $T_{\rm eff}$ (K) | $\log g$ (cm/s$^2$) | $Y_{\rm He}$ | $R$ ($R_\odot$) | $\log L$ ($L_\odot$) | $M$ ($M_\odot$) | $\dot{M}$ ($10^{-6} M_\odot$/yr) | $v_\infty$ (km/s) | $f_{\rm cl}$ K06 | $f_{\rm cl}$ new | parameter source: stellar par. | wind par. | clumping | consistent? |
|---|---|---|---|---|---|---|---|---|---|---|---|---|---|---|
| NGC 2392 | 44000 | 3.6 | .23 | 2.4 | 4.36 | 0.84 | 0.050 | 400 | 1 | | K06 | K06 | K06 | no |
| | ″ | ″ | ″ | ″ | ″ | ″ | 0.471 | 400 | 1 | | ″ | this work | ″ | yes |
| | 40000 | 3.7 | .25 | 1.5 | 3.78 | 0.41 | 0.018 | 400 | 1 | | P04 | P04 | ″ | **yes** |
| | ″ | ″ | ″ | ″ | ″ | ″ | ″ | ″ | | 8 | ″ | ″ | this work | ″ |
| IC 4637 | 52000 | 4.2 | .09 | 1.0 | 3.86 | 0.58 | 0.014 | 1500 | 4 | | K06 | K06 | K06 | no |
| | ″ | ″ | ″ | ″ | ″ | ″ | 0.042 | 700 | 4 | | ″ | this work | ″ | yes |
| | 55000 | 4.6 | .11 | 0.8 | 3.74 | 0.87 | 0.018 | 1500 | | 4 | P04 | P04 | ″ | **yes** |
| NGC 3242 | 75000 | 4.8 | .09 | 0.5 | 3.89 | 0.63 | 0.008 | 2300 | 4 | | K06 | K06 | K06 | yes |
| | 75000 | 5.15 | .12 | 0.3 | 3.50 | 0.53 | 0.004 | 2400 | | 4 | P04 | P04 | ″ | **yes** |
| IC 4593 | 40000 | 3.6 | .09 | 2.2 | 4.12 | 0.70 | 0.042 | 900 | 4 | | K06 | K06 | K06 | no |
| | ″ | ″ | ″ | ″ | ″ | ″ | 0.120 | 250 | 4 | | ″ | this work | ″ | yes |
| | 40000 | 3.8 | .12 | 2.2 | 4.08 | 1.12 | 0.061 | 850 | | 4 | P04 | P04 | ″ | **yes** |
| IC 418 | 36000 | 3.2 | .17 | 4.0 | 4.50 | 0.92 | 0.035 | 700 | 50 | | K06 | K06 | K06 | no |
| | ″ | ″ | ″ | ″ | ″ | ″ | 0.639 | 200 | 50 | | ″ | this work | ″ | yes |
| | 39000 | 3.7 | .1 | 2.7 | 4.22 | 1.33 | 0.098 | 800 | 50 | | P04 | P04 | ″ | **yes** |
| | ″ | ″ | ″ | ″ | ″ | ″ | ″ | ″ | | 4 | ″ | ″ | this work | ″ |
| He 2-108 | 34000 | 3.4 | .09 | 2.6 | 3.99 | 0.62 | 0.136 | 700 | 1 | | K06 | K06 | K06 | no |
| | ″ | ″ | ″ | ″ | ″ | ″ | 0.404 | 300 | 1 | | ″ | this work | ″ | yes |
| | 39000 | 3.7 | .1 | 2.7 | 4.18 | 1.33 | 0.098 | 800 | 1 | | P04 | P04 | ″ | **yes** |
| | ″ | ″ | ″ | ″ | ″ | ″ | ″ | ″ | | 8 | ″ | ″ | this work | ″ |
| Tc 1 | 34000 | 3.2 | .09 | 3.8 | 4.34 | 0.84 | 0.036 | 950 | 10 | | K06 | K06 | K06 | no |
| | ″ | ″ | ″ | ″ | ″ | ″ | 0.605 | 300 | 10 | | ″ | this work | ″ | yes |
| | 35000 | 3.6 | .15 | 3.0 | 4.12 | 1.37 | 0.054 | 850 | 10 | | P04 | P04 | ″ | **yes** |
| | ″ | ″ | ″ | ″ | ″ | ″ | ″ | ″ | | 4 | ″ | ″ | this work | ″ |
| He 2-131 | 32000 | 3.2 | .33 | 3.5 | 4.13 | 0.71 | 0.131 | 400 | 8 | | K06 | K06 | K06 | no |
| | ″ | ″ | ″ | ″ | ″ | ″ | 0.105 | 300 | 8 | | ″ | this work | ″ | yes |
| | 33000 | 3.1 | .30 | 5.5 | 4.59 | 1.39 | 0.321 | 400 | 8 | | P04 | P04 | ″ | **yes** |
| | ″ | ″ | ″ | ″ | ″ | ″ | ″ | ″ | | 4 | ″ | ″ | this work | ″ |
| NGC 6826 | 46000 | 3.8 | .09 | 1.8 | 4.19 | 0.74 | 0.076 | 1200 | 4 | | K06 | K06 | K06 | no |
| | ″ | ″ | ″ | ″ | ″ | ″ | 0.187 | 600 | 4 | | ″ | this work | ″ | yes |
| | 44000 | 3.9 | .13 | 2.2 | 4.25 | 1.40 | 0.178 | 1200 | | 4 | P04 | P04 | ″ | **yes** |

For each CSPN we show the results from three different model runs: P04: stellar and (consistent) wind parameters from Pauldrach et al. (2004); K06f: stellar parameters from Kudritzki et al. (2006), artificially enforcing their wind parameters in our model runs; and K06c: stellar parameters from Kudritzki et al. (2006) and wind parameters consistent with those stellar parameters, as determined by our hydrodynamic calculations of the radiative driving force. For those objects where Kudritzki et al. (2006) have given a clumping factor $f_{\rm cl} \neq 1$ we show all line profiles both including and excluding clumping. Additionally we present redetermined clumping factors which have as usual been derived from a comparison of the observed optical lines to our predicted optical lines obtained from the P04 models.

**NGC 2392.** The central star of NGC 2392 is of particular interest as it has a mass of only 0.41 $M_\odot$, making it the least massive CSPN of our sample. A clumping factor of 1 was derived by Kudritzki et al. (2006), corresponding to an unclumped wind. Unfortunately we lack observations for many of the optical lines for which we show synthetic line profiles in Fig. 9. The computed line profiles, both in absorption and emission, for the consistent P04 and the fitted K06f models are almost identical. Both models fail to reproduce the strong H$\alpha$ and He II $\lambda$ 4686 emission lines but match the absorption lines equally well. The consistent K06c model has a much too high mass loss rate, reflected in a too strong emission of *all* emission lines, and even showing H$\gamma$ in emission. Table 3 shows this huge discrepancy in the mass loss rate which is higher by a factor of 10 in the K06c model compared to the K06f model.

As the consistent P04 and the fitted K06f model both fail to reproduce the strong H$\alpha$ and He II $\lambda$ 4686 emission lines, we have redetermined the clumping factor for the P04 model from a comparison of the observed optical lines to our predictions. We obtain a clumping factor of 8, a strong deviation to the result of Kudritzki et al. (2006) (cf. Table 3). The derived clumping factor represents a compromise between the predictions for the He II $\lambda$ 4686 line and the H$\alpha$ line: a higher clumping factor could have reproduced the observed height of the He II $\lambda$ 4686 line better, but then the width of the predicted H$\alpha$ line would have exceeded that of the observed profile (dash-dotted red line in Fig. 9). This is a clear indication that the physical effect that enhances the emission in these two lines and which we model with the clumping approach does not quite behave, in particular with regard to its radial run, as the simplistic description applied at the present stage.

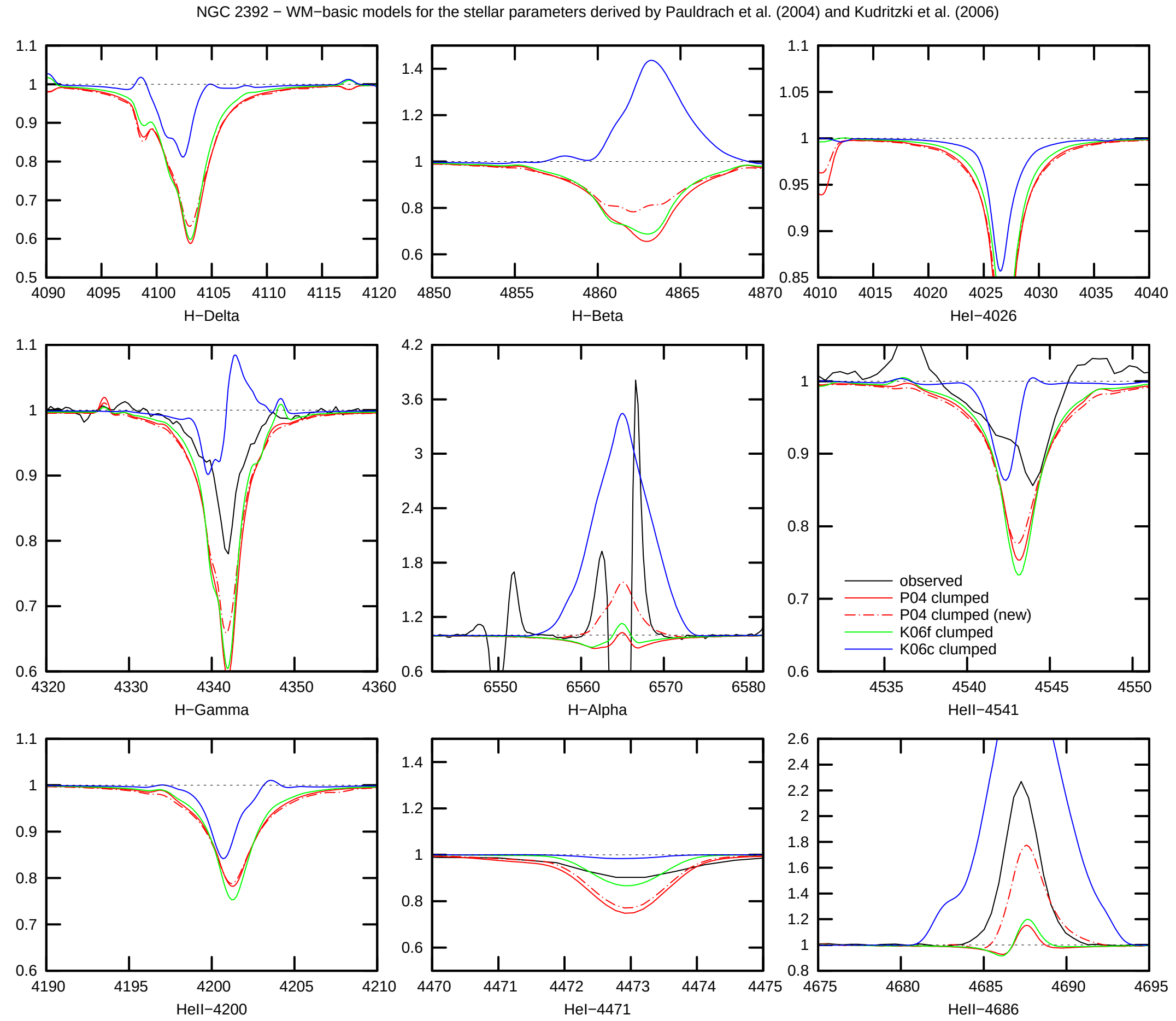

**Fig. 9.** Observed optical line profiles of NGC 2392 (black) compared to predicted line profiles from four different models. Colors indicate different stellar and wind parameter sets, while dash styles indicate different clumping factors. The colors correspond to: the consistent model of Pauldrach et al. (2004) (P04, red); a model using the stellar parameters of Kudritzki et al. (2006) with mass loss rate and terminal velocity artificially set to the values given by Kudritzki et al. (K06f, green); and one using the stellar parameters of Kudritzki et al. (2006) and wind parameters consistent with those stellar parameters (K06c, blue). Solid lines indicate models using the clumping factor suggested by Kudritzki et al. (2006), the dash-dotted line indicates a model using the clumping factor redetermined in this work.

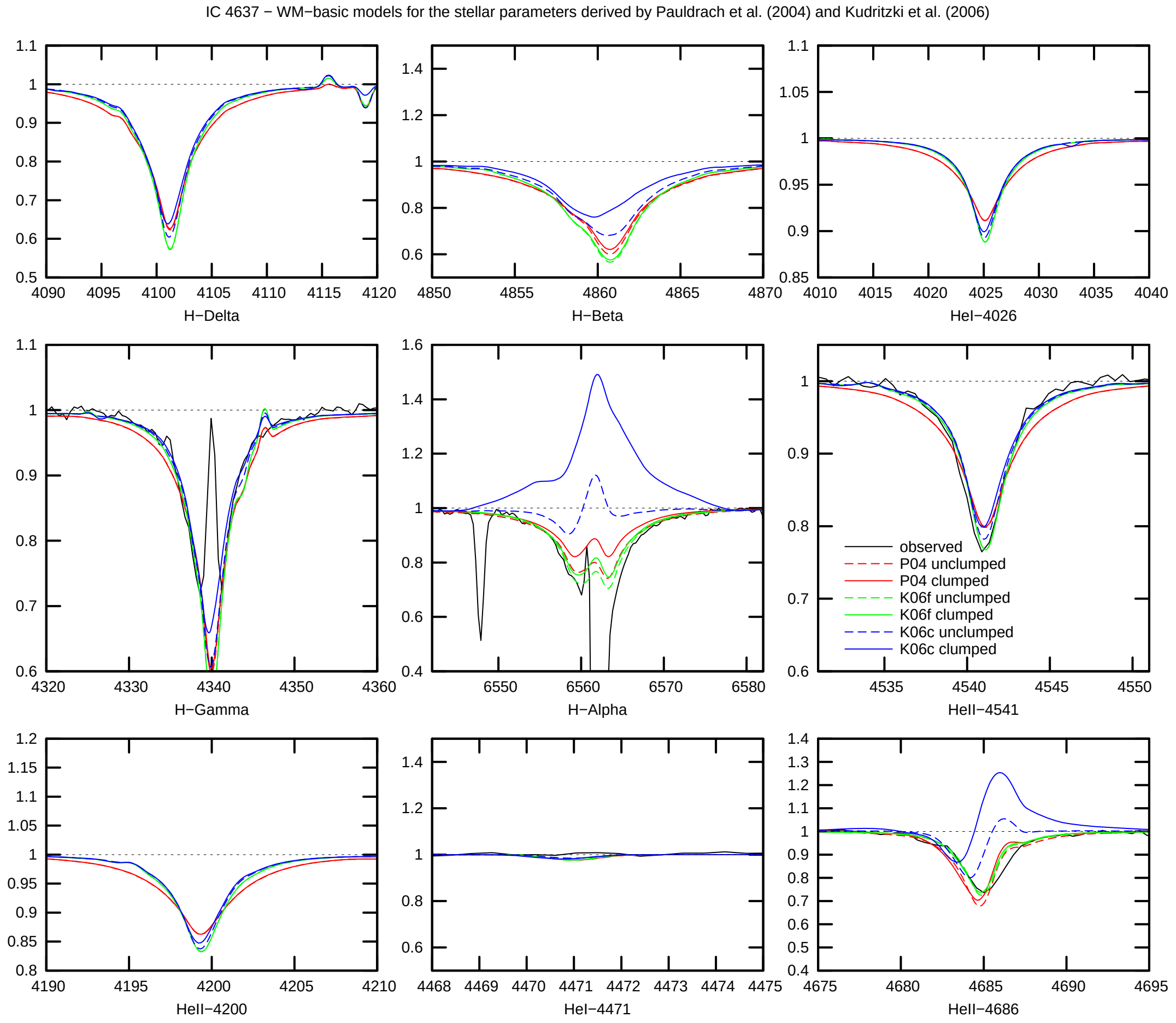

**Fig. 10.** As Fig. 9, but for IC 4637. Profiles from unclumped models are additionally shown dashed.

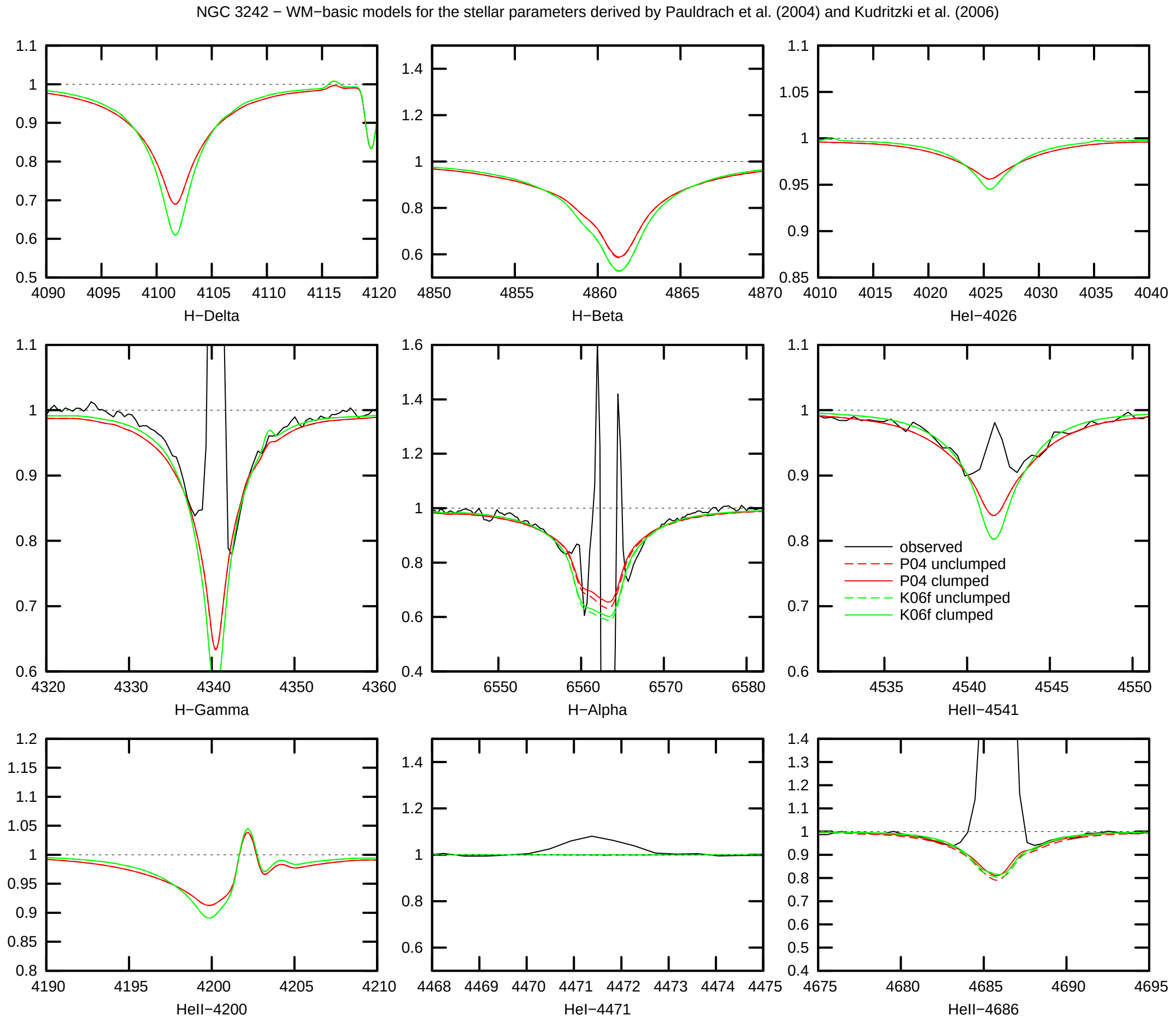

**Fig. 11.** As Fig. 10, but for NGC 3242. This is the only CSPN for which the consistent wind calculation yields the same mass loss rate and terminal velocity as the optical analysis by Kudritzki et al. (2006) and the "green" model is thus identical to the "blue" one (which is therefore not shown).

**IC 4637.** For IC 4637 a clumping factor of 4 was derived by Kudritzki et al. (2006). The resulting synthetic optical line profiles from the P04 and the (inconsistent) K06f models match the observations for IC 4637 (Fig. 10), and both clumped and unclumped wind result in nearly the same profiles (for H$\alpha$, the unclumped P04 model even shows a better fit than the clumped one). The consistent K06c model shows a mostly equally good fit to the absorption lines, but its high emission in H$\alpha$ and He II $\lambda$ 4686 makes the K06 stellar parameters incompatible with the observations.

**NGC 3242.** NGC 3242 is the only CSPN of the sample for which the UV analysis of Pauldrach et al. (2004) and the optical analysis of Kudritzki et al. (2006) yield almost the same mass, and the K06f parameter set is consistent (and thus identical to the K06c model). Both the P04 and the K06f/K06c models give nearly identical fits to all observed optical line profiles (Fig. 11). The effects of clumping are marginal, as neither H$\alpha$ nor He II $\lambda$ 4686 are in emission.

**IC 4593.** For IC 4593 the situation is somewhat similar to that of IC 4637: the observed optical absorption lines are matched equally well by the P04 and the (inconsistent) K06f models (Fig. 12). The predicted emission in H$\alpha$ and He II $\lambda$ 4686 is a bit too low for both models, but with some cosmetics, i.e., a slightly larger clumping factor, the P04 model would show a perfect fit to both lines. The consistent K06c model again produces too strong emission, even without clumping, and thus rules out the K06 stellar parameters.

**IC 418.** IC 418 is of special interest because Kudritzki et al. (2006) have found a clumping factor of 50 (!) for this star, which is by far the highest value, but even with this large clumping factor the K06f model fails to reproduce the observed H$\alpha$ profile (Fig. 13). The consistent model K06c produces much too strong emission and thus gives incompatible line profiles for *all* observed lines, with and without clumping. (Additionally, it results in a terminal velocity of only 200 km/s, which is much too small compared to the observed terminal velocity of 700 km/s.) For the P04 model, this clumping factor is also much too large, but a moderate clumping factor of 4 yields with respect to a compromise of the strength and width of the emission lines of H$\alpha$ and He II $\lambda$ 4686 the best fit to all lines (Fig. 13).

**He 2-108.** Kudritzki et al. (2006) found no evidence of clumping in the wind of He 2-108, but with our computed density structure their mass loss rate does not yield enough emission in H$\alpha$ and He II $\lambda$ 4686 (a situation somewhat similar to NGC 2392). A moderate clumping factor of 8, again a (slight) compromise between width and height of the profiles, yields a much improved fit (Fig. 14), but the same caveat from the discussion of NGC 2392 with regard to the radial run of the clumping factor applies here as well.

He 2-108 is one of the few CSPNs for which the K06c model yields fits to the optical line strength which are of the same quality as the K06f model. However, the K06c model produces a terminal velocity of only 300 km/s, which is too small compared to the observed value.

**Tc 1.** For Tc 1 Kudritzki et al. (2006) found a clumping factor of 10. In the (inconsistent) K06f model, this clump-

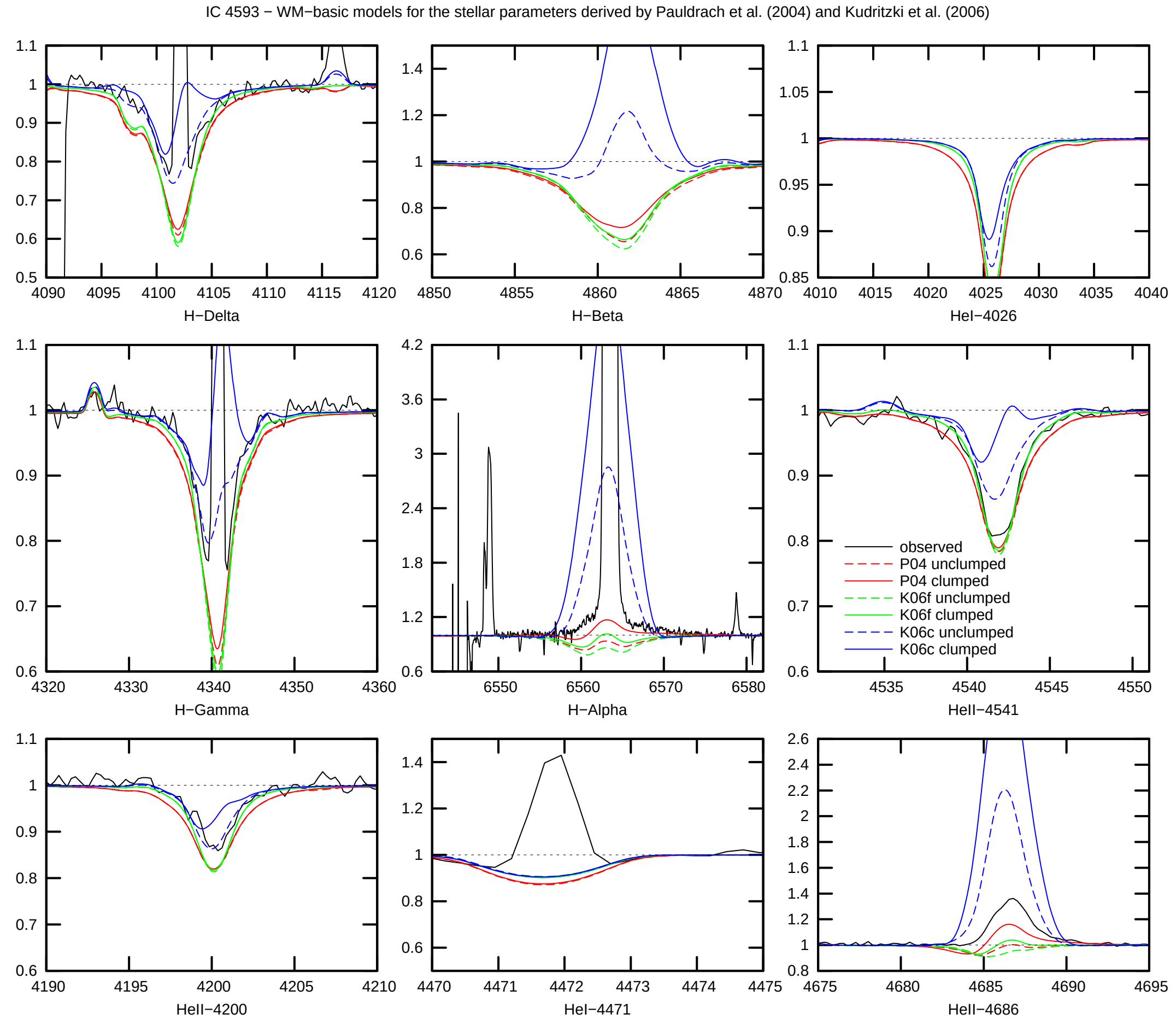

**Fig. 12.** As Fig. 10, but for IC 4593.

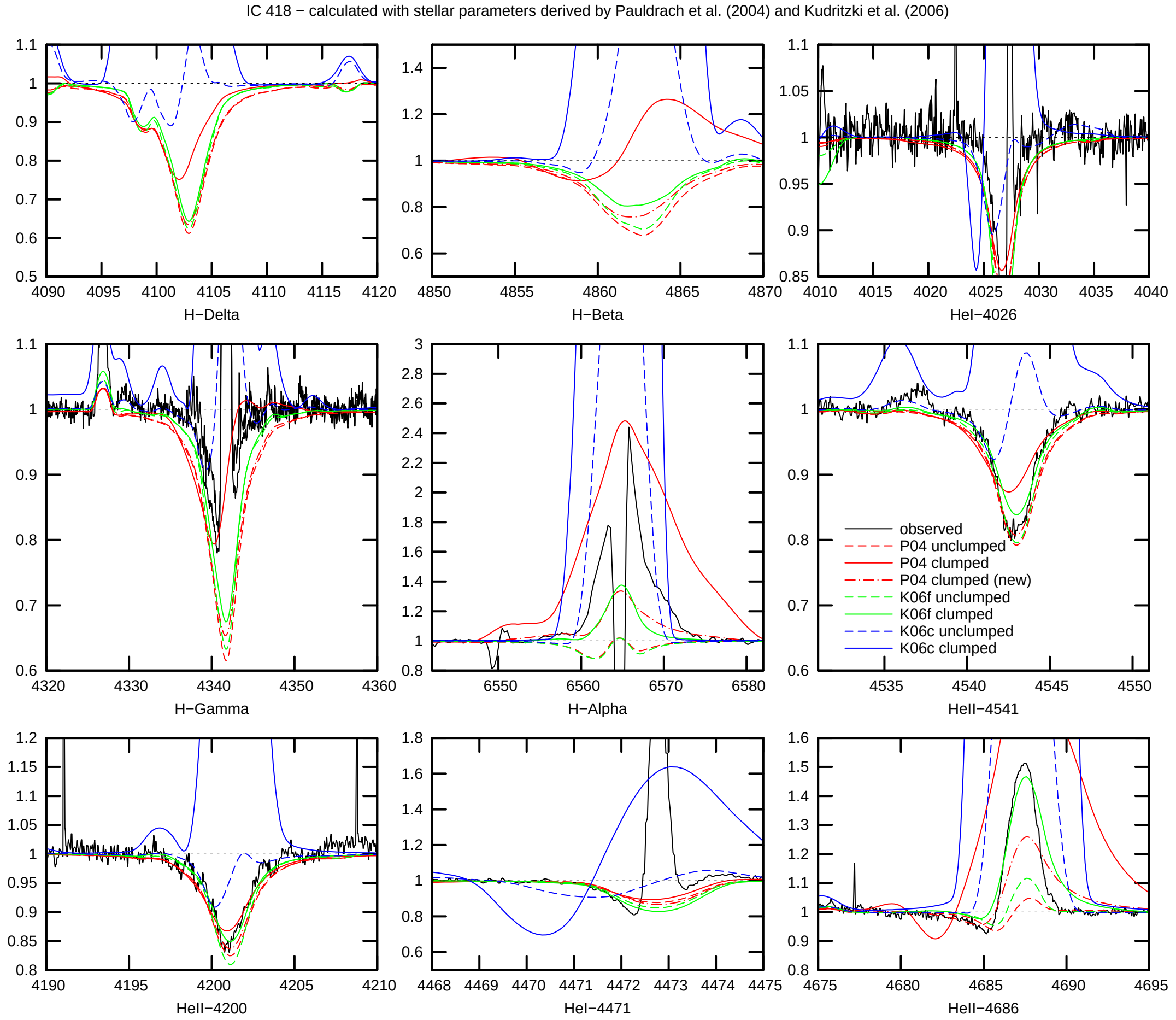

**Fig. 13.** As Fig. 10, but for IC 418.

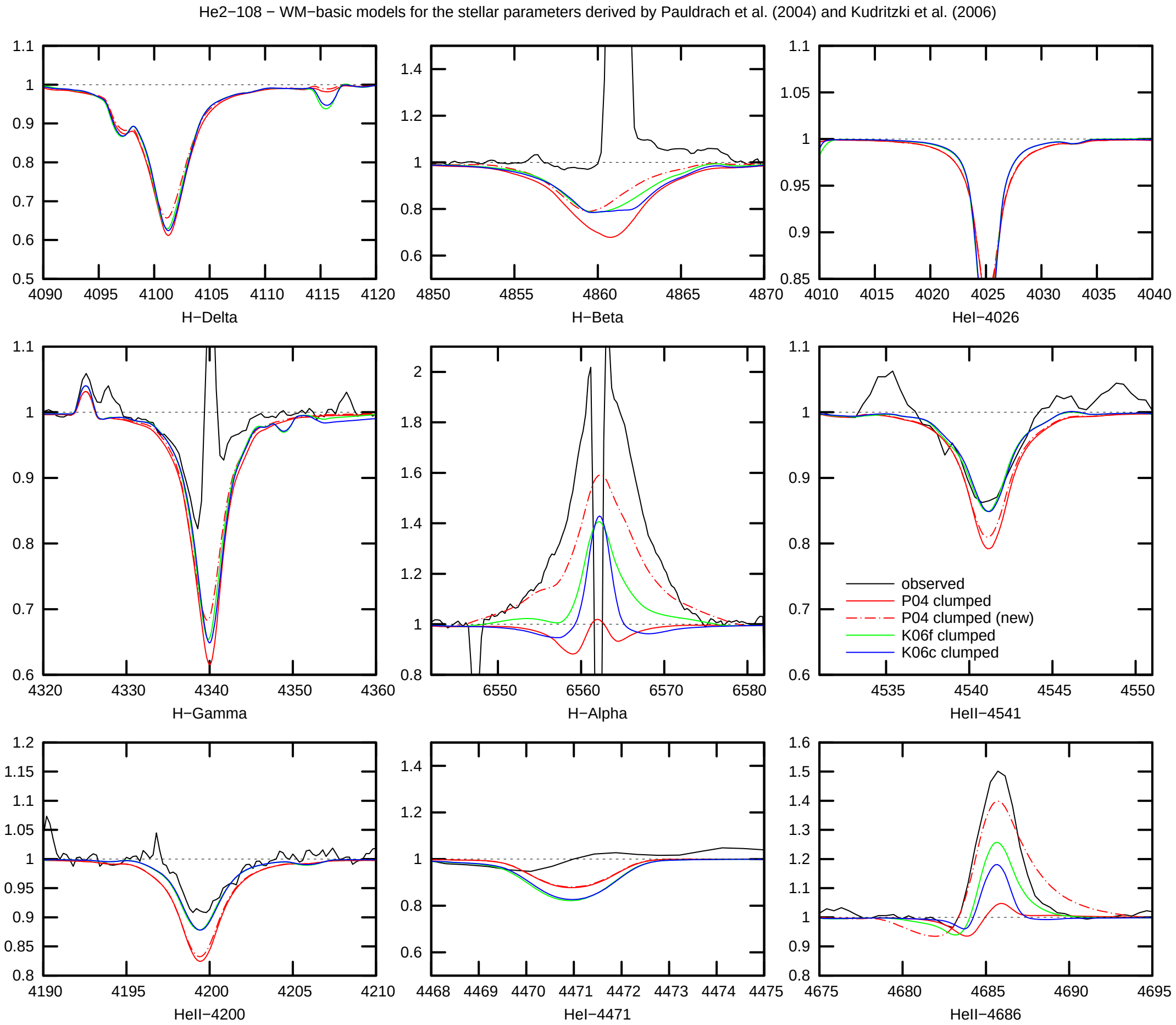

**Fig. 14.** As Fig. 9, but for He 2-108.

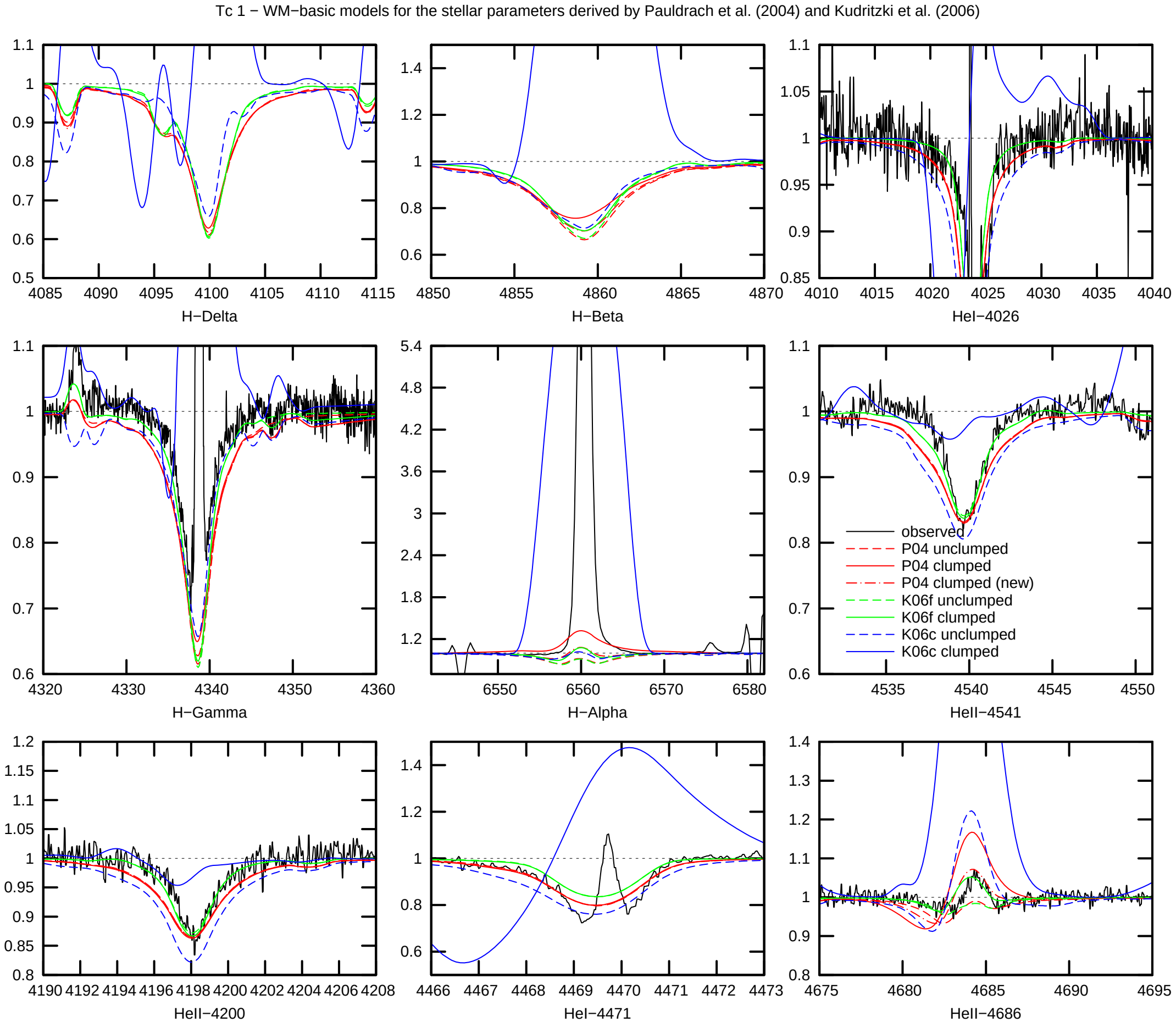

**Fig. 15.** As Fig. 10, but for Tc 1.

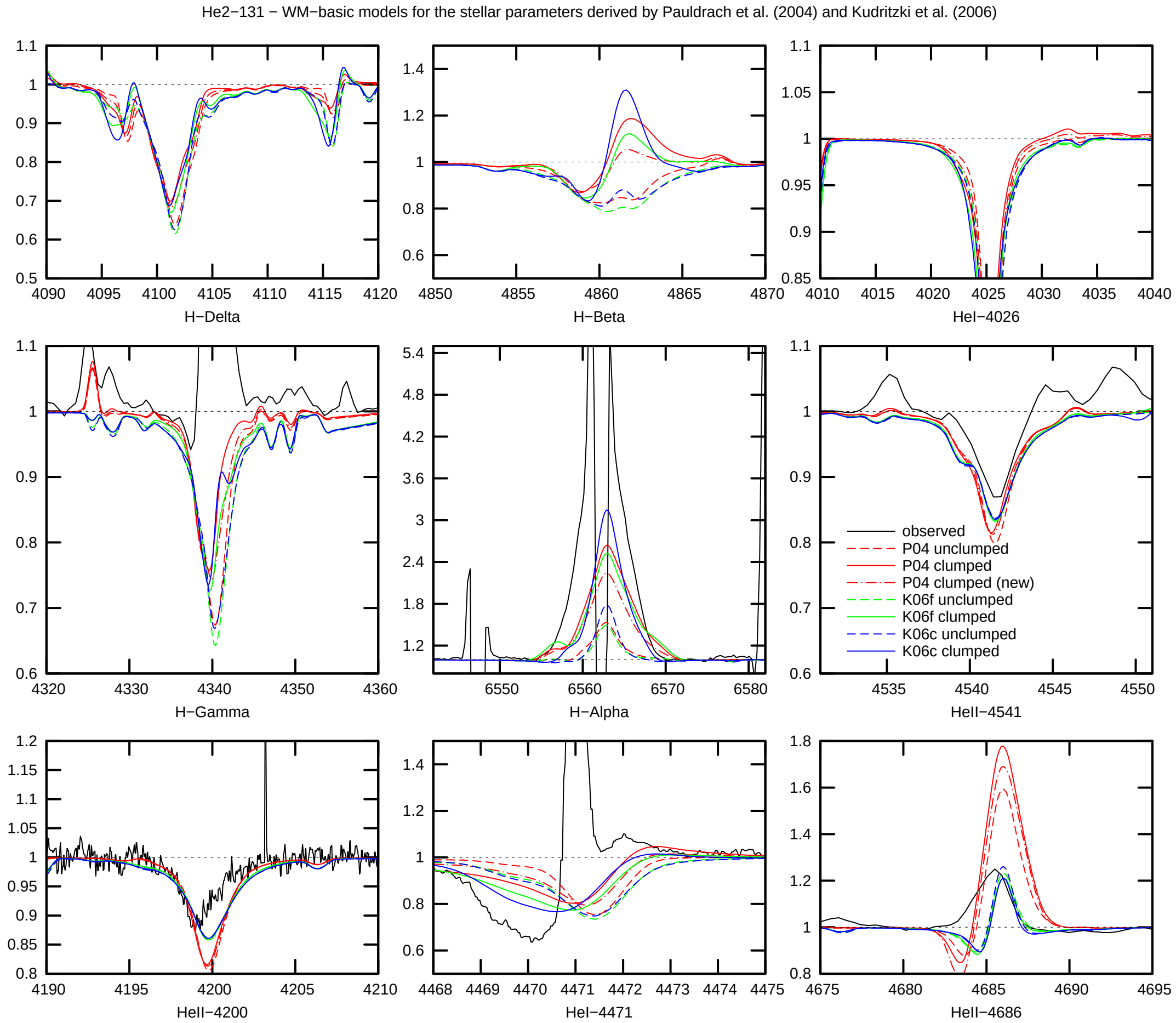

**Fig. 16.** As Fig. 10, but for He 2-131.

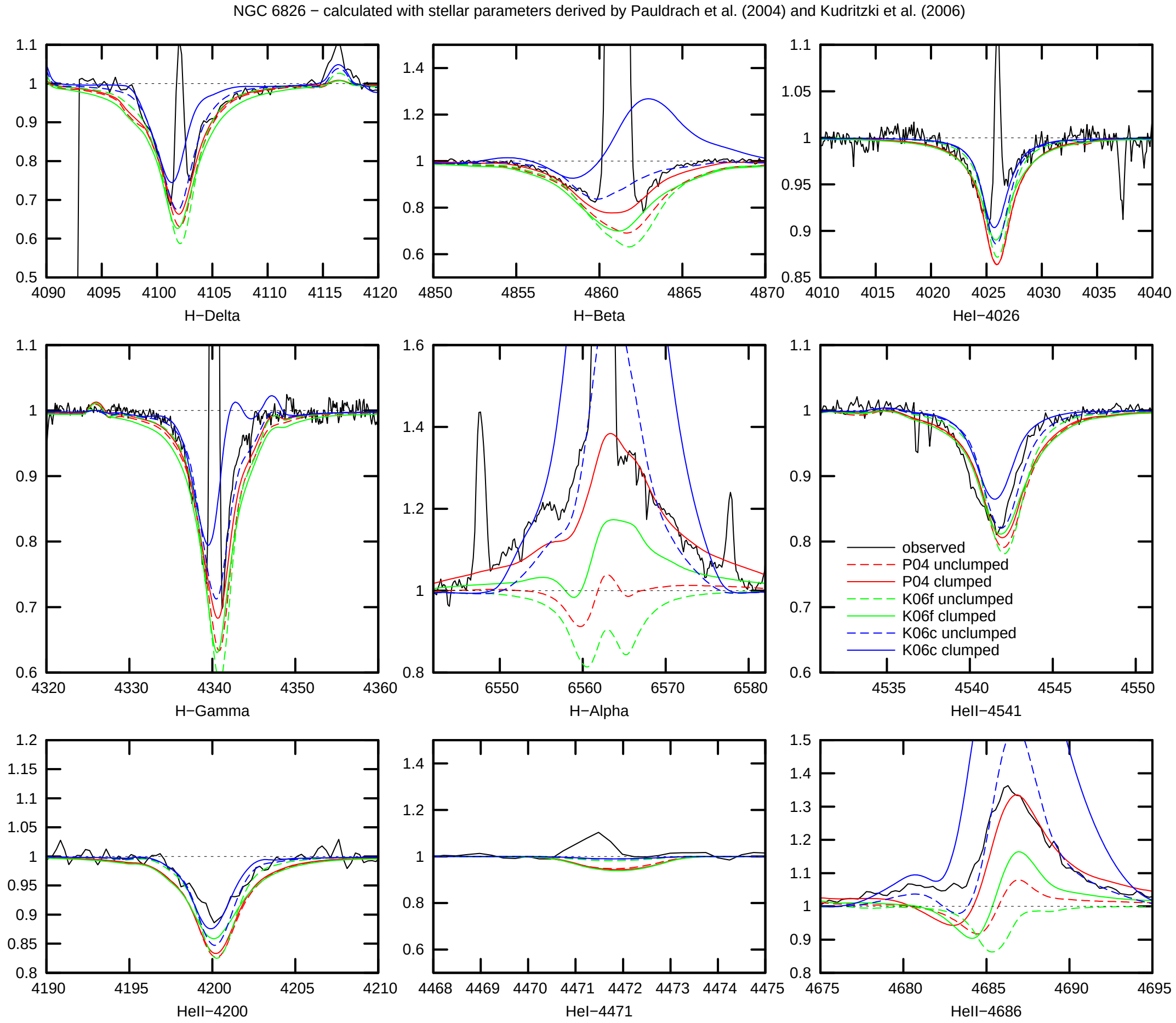

**Fig. 17.** As Fig. 10, but for NGC 6826.

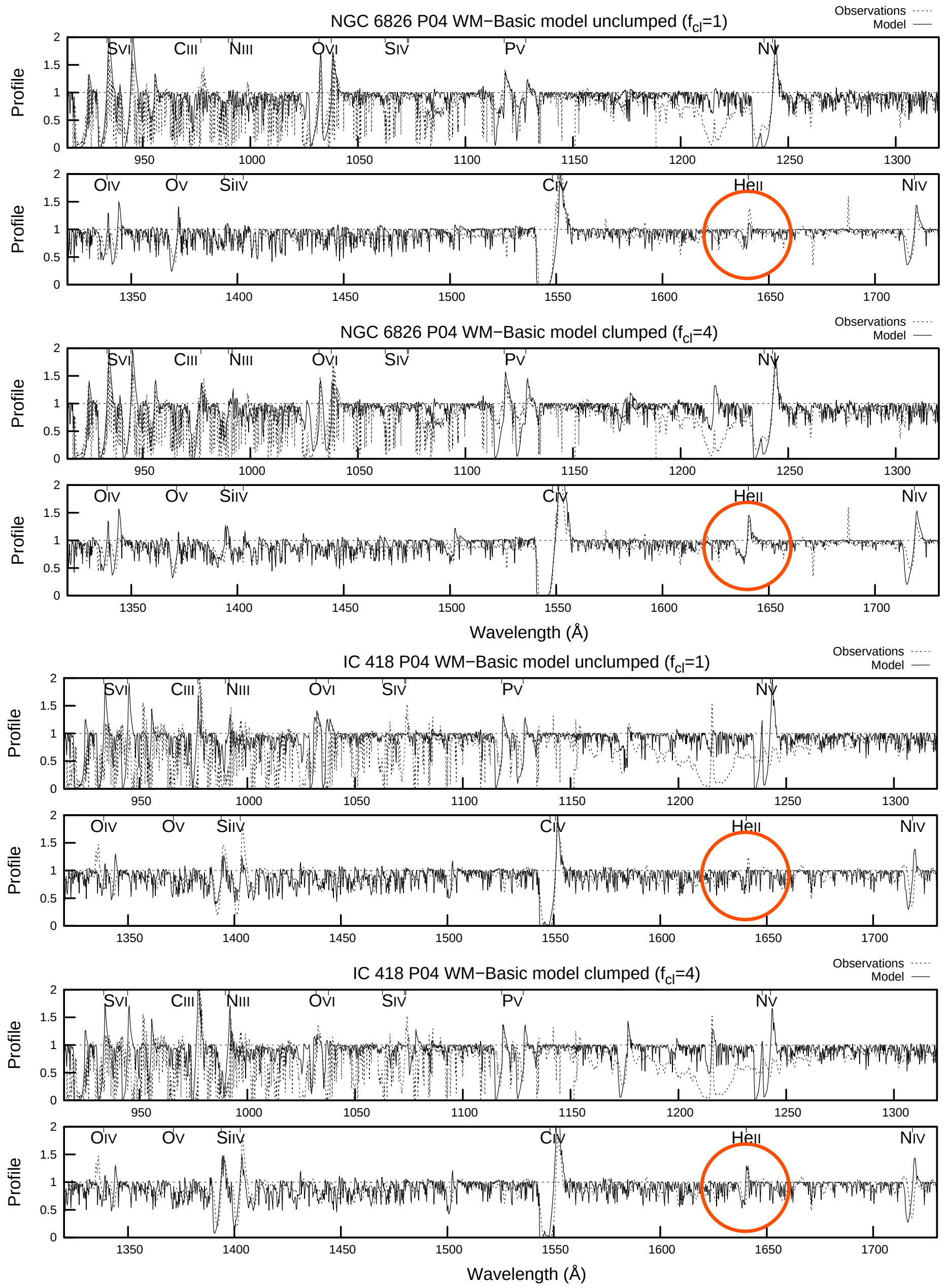

**Fig. 18.** Synthetic UV spectra of NGC 6826 (upper two panels) and IC 418 (lower two panels) calculated with the stellar parameters determined by the UV analysis, compared to the observed UV spectra. The first panels in the upper and lower part show unclumped models. The overall fit to the observed UV spectra is generally very good, although the He II $\lambda$ 1640 line is not reproduced too well. The second panels in the upper and lower parts show the same models, but with a clumping factor of $f_{\rm cl} = 4$. Most of the UV spectrum is not affected, except for a few lines from subordinate ionization stages (such as P V $\lambda$ 1118, C III $\lambda$ 977, and H I $\lambda$ 1216). In particular the match to the observed He II $\lambda$ 1640 line is now, in a cosmetic sense, almost excellent.

ing factor yields a good fit to He II $\lambda$ 4686, but the emission in H$\alpha$ is still slightly low. The P04 model gives too much emission in both H$\alpha$ and He II $\lambda$ 4686 with this clumping factor, and the somewhat lower value of $f_{\rm cl} = 4$ yields a much better fit (Fig. 15). The consistent K06c model has already too much emission in He II $\lambda$ 4686 without clumping, and its terminal velocity of 300 km/s is also too small compared to the observed value of about 900 km/s.

**He 2-131.** For He 2-131 the consistent calculation (K06c) using the stellar parameters of Kudritzki et al. (2006) yields wind parameters similar to those used by Kudritzki et al. (K06f), and their predicted line profiles are thus also very similar (Fig. 16). All three model calculations yield similar profiles with too little emission in H$\alpha$ even with a clumping factor of 8, and also do not correctly reproduce He II $\lambda$ 4686, P04 by yielding too much emission and K06c and K06f too little. The H$\alpha$ line again shows tendencies of becoming too wide with this clumping factor, another indication that clumping should be reduced at higher outflow velocities, and a global clumping factor of 4 does not significantly change the appearance (Fig. 16). The P04 model, using the helium abundance determined by Kudritzki et al. (1997), shows too much emission already without clumping, indicating that the rather high helium abundance assumed might in reality be smaller, more like that of the other stars of the sample.

The absorption lines are of similar quality in all three models, and are fairly unaffected by our cosmetic procedure.

**NGC 6826.** NGC 6826 is the CSPN with the highest mass of the whole sample: for this star Pauldrach et al. (2004) derived a mass of 1.40 $M_\odot$, close to the Chandrasekhar limit for white dwarfs, thus making this star a possible type Ia supernova progenitor. With the clumping factor of $f_{\rm cl} = 4$ suggested by Kudritzki et al. (2006) the P04 model yields acceptable fits to *all* lines, including the emission lines H$\alpha$ and He II $\lambda$ 4686. This result is interesting as the *fitted* K06f model does not reproduce the strength of these emission lines (Fig. 17) with our computed density structure, yielding too little emission. For the absorption lines all models give equally good fits, except for the clumped K06c model, which produces too much emission in several absorption lines.

**Clumping and the He II $\lambda$ 1640 line.** A number of CSPNs from our sample have a pronounced He II $\lambda$ 1640 profile in the UV spectrum, which may be used to cross-check the diagnosed clumping factor from the optical He II $\lambda$ 4686 line. In Fig. 18 we compare the observed UV spectra of the two CSPNs NGC 6826 and IC 418 to corresponding synthetic spectra from models including clumping. As we had already seen from the optical He II $\lambda$ 4686 line profile of IC 418 that a clumping factor of $f_{cl} = 50$ as suggested by Kudritzki et al. (2006) was too large to be compatible with the observations of this star, we have used a smaller clumping factor of $f_{cl} = 4$ for the IC 418 model we show here.

The upper panel of each figure shows the synthetic spectrum from the unclumped model, the lower panel the synthetic spectrum from a model with identical parameters, but with clumping applied. As the plots clearly show, *moderate clumping does not influence the generally good overall fit of the predicted spectra. In fact, most of the UV spectrum is unaffected by clumping*, except for a few lines from subordinate ionization stages (such as P V, C III, and H I). In particular, clumping also enhances the occupation of He II (the main ionization stage of helium in the photosphere and most of the wind is He III) and leads to an improved fit of the He II $\lambda$ 1640 profile. However, the quality of the fit of this line is of no importance for judging the fundamental model parameters; and this is also the case for the emission lines of H$\alpha$ and He II $\lambda$ 4686 discussed above.

But, with regard to our finding that the UV spectrum of our stars is barely influenced by moderate clumping reflecting the fact that the main ionization stages of the elements are not severely affected, we can nevertheless draw an important conclusion from this investigation: As the increased recombination in the denser clumps leads to a higher occupation of mostly the ionization stages below the main ionization stage, which usually have much smaller occupation numbers than the main ionization stage, and as the main ionization stages, and not the subordinate ones, are the primary contributors to the line force, moderate clumping also has little influence on the hydrodynamics of the outflow, and thus on the consistency of the Pauldrach et al. (2004) stellar and wind parameters.

## 6. Discussion, summary, and conclusions

The central point of the analysis presented here for a selected sample of CSPNs is related to a correct treatment of line-driving – the mechanism that determines the strength of the outflow and the shape of the velocity field, which in turn have a strong influence on the appearance of the spectra. Although it is generally accepted that the winds of O-type stars have this driving mechanism, it appears that the majority is not aware of what the observed properties of the winds, together with the behavior of this driving mechanism, imply for the stellar parameters of these objects. And this has nothing to do with the details of spectral modelling, only with the fact that line-driving predicts the mass loss rate and the terminal velocity to be monotonous functions of each of the stellar parameters. The aim of this paper has been to show that all of the observed behavior of O star winds is perfectly understandable as being driven by radiation pressure. This is in particular also true for the stars of our CSPN sample, if one accepts that these selected objects do not follow the theoretical mass–luminosity relation for post-AGB stars.

The approach to spectral analysis of O-type atmospheres with winds used by many others in the field, simply postulating a velocity law, adapting the main parameter of that law to get the best overall reproduction of observed optical emission profiles, and then attributing everything which cannot be fitted to “clumping”, is in reality not a method that can be used to derive stellar masses and radii, because it cannot predict the behavior of the winds (and thus, of the spectra) as a function of the stellar parameters. Since the observed features are modelled using free parameters without any physical constraints of cause and effect, such a method has no predictive power and cannot actually explain anything about the physics of the atmosphere. Although the ad-hoc implementation of clumping in such methods has given the illusion of being able to understand O-star winds, the results achieved may be misleading because they are possibly based on wrong premises. Without the consistency between wind and stellar parameters the treatment of secondary effects such as clumping, while certainly established for some years, is just decoration, because it does not address the fundamental issue. (The approach is similar to introducing epicycles into the geocentric world model, which, despite the epicycles, remains a fundamentally misguided theory.) All work in which the primary influence on the line strengths is treated on the basis of pure guesswork – an assumed velocity field and a mass loss rate that has not been shown to be consistent – must be considered to be ultimately flawed.

In this regard the most important caveats raised by this paper are: The differences between the assumed and the computed velocity fields – for the same terminal velocity and mass loss rate – lead to an effect on the line profiles on the same scale as the assumed clumping. Thus, it makes no sense at all to try to pin down clumping as long as a more fundamental influence – that of the basic structure of the velocity field – is not properly considered. And the mass loss rate (and the terminal velocity) are not free parameters that one may arbitrarily adjust to fit the spectrum. Instead, a set of stellar parameters must be found that can actually drive a wind with which the observed spectrum can be reproduced in a consistent way. Thus, it makes even less sense to try to pin down clumping as long as the most fundamental influence – *consistent* mass loss rate and terminal velocity – is not properly considered.

On the basis of these central points we have presented, using the two mutually contradicting sets of stellar parameters published in the literature for a sample of CSPNs, new comparisons of observed and model spectra in order to provide further clues to the true parameters of the stars of this sample. Our comparisons not only include the extended spectral range of 905 to 1085 Å observed by FUSE in addition to the IUE spectral range of 1150 to 1980 Å and the optical lines, but also consider the available X-ray observations. They also include a treatment of wind clumping in the models, which in the literature has been considered to be essential to reproduce the observed optical emission lines.

The FUSE spectral range includes the wavelengths of the O VI resonance doublet. As O VI is known from massive O stars to be primarily produced by Auger ionization driven by X-rays originating from the cooling zones of shocks resulting from the unstable, non-stationary behavior of the winds, this line therefore represents a special diagnostic tool for wind-embedded shocks. The failure to reproduce the observed O VI line in the synthetic spectra of standard CSPN models strongly indicates that X-rays are an important ingredient in CSPN winds, and this information is also provided by X-ray observations that show O-type central stars of planetary nebulae as X-ray emitters. By ap-plying the physics which describe this phenomenon in our CSPN simulations we could show that the synthetic spectra of our current best models reproduced the profile shapes of the O VI P-Cygni line quite well. As an integral part of the modelling procedure we obtained the emitted frequency-integrated relative X-ray luminosities as well as the radial distribution of the maximum local shock temperatures in the cooling zones of the shock-heated matter component.

In a second step these values have been compared with those resulting directly from observations. 50% of the planetary nebulae observed by a recent Chandra survey are X-ray emitters, with both diffuse and point-like X-ray sources. A detailed measurement of the X-ray emission of one of the objects (NGC 2392) in our sample exists, which displays both a diffuse and a point-like component. The total measured X-ray emission is about a factor of two larger than the value we have deduced from our best-fit model for this CSPN. Thus our result is not only in accordance with the observations, but also with the interpretation of these observations. Moreover, the maximum shock temperatures we have derived are also in agreement with these current observations. The ratios of X-ray to bolometric luminosity we determined for the stars of our sample ($L_\mathrm{X}/L_\mathrm{bol} \sim 10^{-7} \dots 10^{-6}$) are also typical for massive O stars, and are further confirmed by another observation which obtained a similar value for another CSPN not in our sample.

These similarities between massive O stars and O-type CSPNs justify using the same tools that had been applied successfully to the analysis of massive O star spectra also to O-type CSPNs. Indeed, the stellar parameters that had been determined from the UV analysis (Pauldrach et al. 2004) offer a consistent picture not only of the X-ray, the UV, and the optical spectral properties of CSPNs, but also of the wind-dynamical properties not directly related to the spectral appearance, such as the ratios $v_\infty/v_\mathrm{esc}$ of terminal wind velocity to stellar escape velocity (Kaschinski et al. 2012) and the agreement of the wind strengths with the wind-momentum–luminosity relation (WMLR).

This is in contrast to the stellar parameters that had been determined from the optical line profiles with the aid of the theoretical mass–luminosity relation of CSPNs predicted by current post-AGB evolutionary models (Kudritzki et al. 2006). The spectra from models using these parameters and consistently calculated wind dynamics reproduce the observations not at all. Furthermore, for a number of stars in the sample these stellar parameters lead to anomalously high $v_\infty/v_\mathrm{esc}$ ratios. From this we conclude that the published optical analyses give good fits to the observed spectra only because the wind parameters assumed in these analyses were inconsistent to their stellar parameters.

Although a complete set of stellar parameters can be determined for O-type central stars with expanding atmospheres just from the wind-sensitive features in the UV, a comparison which also includes the H and He lines of the optical spectra has, as a combined analysis, a more convincing character. To establish such a procedure, we had already improved our stellar atmosphere code WM-basic by including Stark broadening in order to describe the optical lines in a more realistic way. A still open issue, however, were the fits to the optical recombination lines H$\alpha$ and He II $\lambda$4686, whose emission could not be fully reproduced. As the current consensus appears to be that these lines are influenced by clumping in the winds, we reinvestigated these lines via a comparison of clumped and unclumped models for our chosen CSPN sample.

For those CSPNs with emission lines in the optical, moderate clumping factors increase the emission and provide good fits to the observed line profiles when using the parameters of Pauldrach et al. (2004), thus supporting their parameter determination. The stellar parameters of Kudritzki et al. (2006) together with consistent wind parameters, on the other hand, gave line profiles that were not compatible with the observed line profiles.

When artificially forcing our models to use the mass loss rates of Kudritzki et al. (2006) with their stellar parameters, the fits were better than with the consistent mass loss rates and terminal velocities but not as good as those of Kudritzki et al. (2006). This is because our WM-basic code computes the density structure from a solution to the equation of motion balancing gravity and gas and radiation pressure, instead of simply parametrizing the velocity law. Thus, Kudritzki et al. (2006) could achieve good fits to the optical line profiles only by also assuming an unrealistic density structure that was not compatible with their stellar parameters. A realistic, consistent density and velocity structure with these stellar parameters yields profiles incompatible with the observations and thus rules out these stellar parameters. Thus, lines whose model profiles are determined primarily by a cosmetic clumping factor and not the underlying physics of the model completely lose their diagnostic value, and the quality of the fit of these lines says nothing about the reliability of the fundamental model parameters if the model is already based on an unrealistic density structure.

Our general finding that the observable quantities of the X-ray to bolometric luminosity ratios ($L_\mathrm{X}/L_\mathrm{bol} \sim 5 \times 10^{-7}$) and the maximum shock temperatures ($T^\mathrm{max}_\mathrm{shock} \sim 10^6$ K) have similar values for all CSPNs of the sample is consistent with our finding that the clumping factors are also similar, with a typical value of $f_\mathrm{cl} \sim 4$. This is an important result, since it shows that *the apparently erratic character of the clumping factors seen in the optical analyses vanishes if the calculations are based on consistent sets of stellar and wind parameters*. As an erratic behavior of the clumping factors is not to be expected from our knowledge of shocks arising from the radiative wind driving mechanism, this result supports the legitimate assumption that the clumping factors must be correlated with the other observables of the shock physics.

In this context we also showed that the UV spectrum of our stars is almost unaffected by moderate clumping, and the addition of clumping to the model does not worsen the generally good overall fit of our spectra previously predicted without clumping. Only a few lines from subordinate ionization stages are affected by clumping, among those the subordinate He II $\lambda$1640 line for which we now achieve a “cosmetically good” fit. That in this CSPN sample the UV spectrum is barely influenced by moderate clumping reflects the fact that the main ionization stages of the elements are not severely affected, and thus clumping has only a small influence on the line force and the hydrodynamics of the outflow, and therefore on the consistency of the dynamically determined stellar and wind parameters.

Thus, *second order effects like clumping do not influence the primary results of the spectral analysis* and they have only a marginal influence on the major spectral characteristics – the UV metal lines – of O stars. The self-consistent hydrodynamic approach remains therefore not only the superior approach, but is in fact *essential* for the analysis, since with a model that properly implements the physics of radiation driven winds a good fit to the observations can only be achieved with a proper choice of a complete set of stellar parameters, and in this regard there is simply *no alternative* to a self-consistent hydrodynamical method.

By applying this method we obtained for five out of the nine CSPNs of our sample masses near, but not above, the critical Chandrasekhar mass limit for white dwarfs. Ignoring the percentage number of these objects – since our sample is most prob-ably influenced by selection effects – this result is certainly of relevance for the still controversially discussed precursor scenarios of type Ia supernovae, as these objects might belong to a not yet understood subgroup of CSPNs that evolve to white dwarfs which can end up as supernovae of type Ia.

*Acknowledgements.* We thank Joachim Puls for his advice and for the FASTWIND comparison models; Rolf Kudritzki for making the optical observations available to us; and Miguel Urbaneja for sharing his original CSPN model runs. We also thank the anonymous referee for helpful comments which improved the paper. This work was supported by the Deutsche Forschungsgemeinschaft (DFG) under grants Pa 477/4-1 and Pa 477/7-1.

## References

Castor, J. I., Abbott, D. C., & Klein, R. I. 1975, ApJ, 195, 157
Chandrasekhar, S. 1931, ApJ, 74, 81
Guerrero, M. A. 2006, in IAU Symposium, Vol. 234, Planetary Nebulae in our Galaxy and Beyond, ed. M. J. Barlow & R. H. Méndez, 153–160
Guerrero, M. A., Chu, Y.-H., & Gruendl, R. A. 2000, ApJS, 129, 295
Guerrero, M. A., Chu, Y.-H., Gruendl, R. A., & Meixner, M. 2005, A&A, 430, L69
Herald, J. E. & Bianchi, L. 2011, MNRAS, 417, 2440
Kaschinski, C. B., Pauldrach, A. W. A., & Hoffmann, T. L. 2012, A&A, 542, A45
Kastner, J. H., Montez, Jr., R., Balick, B., et al. 2012, AJ, 144, 58
Kudritzki, R. 1999, in Lecture Notes in Physics, Berlin Springer Verlag, Vol. 523, IAU Colloq. 169: Variable and Non-spherical Stellar Winds in Luminous Hot Stars, ed. B. Wolf, O. Stahl, & A. W. Fullerton, 405
Kudritzki, R. P., Méndez, R. H., Puls, J., & McCarthy, J. K. 1997, in IAU Symposium, Vol. 180, Planetary Nebulae, ed. H. J. Habing & H. J. G. L. M. Lamers, 64
Kudritzki, R. P., Urbaneja, M. A., & Puls, J. 2006, in IAU Symposium, Vol. 234, Planetary Nebulae in our Galaxy and Beyond, ed. M. J. Barlow & R. H. Méndez, 119
Lucy, L. B. 1982, ApJ, 255, 286
Lucy, L. B. & Solomon, P. M. 1970, ApJ, 159, 879
Lucy, L. B. & White, R. L. 1980, ApJ, 241, 300
Méndez, R. H., Kudritzki, R. P., Groth, H. G., Husfeld, D., & Herrero, A. 1988a, A&A, 197, L25
Méndez, R. H., Kudritzki, R. P., Herrero, A., Husfeld, D., & Groth, H. G. 1988b, A&A, 190, 113
Milne, E. A. 1926, MNRAS, 86, 459
Owocki, S. P. & Rybicki, G. B. 1984, ApJ, 284, 337
Pauldrach, A. 1987, A&A, 183, 295
Pauldrach, A., Puls, J., Kudritzki, R.-P., Méndez, R. H., & Heap, S. R. 1988, A&A, 207, 123
Pauldrach, A. W. A., Feldmeier, A., Puls, J., & Kudritzki, R. P. 1994a, in Evolution of Massive Stars, ed. D. Vanbeveren, W. van Rensbergen, & C. De Loore, 105–125
Pauldrach, A. W. A., Hoffmann, T. L., & Lennon, M. 2001, A&A, 375, 161
Pauldrach, A. W. A., Hoffmann, T. L., & Méndez, R. H. 2003, in IAU Symposium, Vol. 209, Planetary Nebulae: Their Evolution and Role in the Universe, ed. S. Kwok, M. Dopita, & R. Sutherland, 177
Pauldrach, A. W. A., Hoffmann, T. L., & Méndez, R. H. 2004, A&A, 419, 1111
Pauldrach, A. W. A., Kudritzki, R.-P., Puls, J., & Butler, K. 1990a, A&A, 228, 125
Pauldrach, A. W. A., Kudritzki, R. P., Puls, J., Butler, K., & Hunsinger, J. 1994b, A&A, 283, 525
Pauldrach, A. W. A. & Puls, J. 1990a, Reviews of Modern Astronomy, 3, 124
Pauldrach, A. W. A. & Puls, J. 1990b, A&A, 237, 409
Pauldrach, A. W. A., Puls, J., Gabler, R., & Gabler, A. 1990b, in Properties of hot luminous stars, Boulder-Munich workshop, ed. C.D. Garmany, Vol. PASPC 7, 171–188
Pauldrach, A. W. A., Vanbeveren, D., & Hoffmann, T. L. 2012, A&A, 538, A75
Prinja, R. K., Massa, D. L., Urbaneja, M. A., & Kudritzki, R.-P. 2012, MNRAS, 422, 3142
Puls, J., Markova, N., Scuderi, S., et al. 2006, A&A, 454, 625
Repolust, T., Puls, J., & Herrero, A. 2004, A&A, 415, 349
Runacres, M. C. & Owocki, S. P. 2002, A&A, 381, 1015
Sana, H., Rauw, G., & Gosset, E. 2005, in Massive Stars and High-Energy Emission in OB Associations, ed. G. Rauw, Y. Nazé, R. Blomme, & E. Gosset, 107–110
Sana, H., Rauw, G., Nazé, Y., Gosset, E., & Vreux, J.-M. 2006, MNRAS, 372, 661
Snow, T. P. & Morton, D. C. 1976, ApJS, 32, 429